\documentclass[altaffilletter,aps,nofootinbib,twocolumn,prd,eqsecnum,preprintnumbers,superscriptaddress,10pt,floatfix]{revtex4-1}
\pdfoutput=1
\usepackage{graphicx}
\usepackage{amsmath}
\usepackage{amssymb}
\usepackage{amsfonts}
\usepackage{mathtools}
\usepackage{amssymb}
\usepackage{enumerate}
\usepackage{xcolor}
\usepackage{bm}
\usepackage{mathrsfs}
\usepackage{epstopdf}
\usepackage{url}
\usepackage{footnote}
\usepackage{textcomp}
\usepackage{dsfont}
\usepackage{ulem}
\usepackage{hyperref}
\usepackage{enumerate}   
\usepackage{appendix}
\usepackage{textcomp}
\usepackage{tipa}

\makeatletter
\newcommand*{\rom}[1]{\expandafter\@slowromancap\romannumeral #1@}
\makeatother
\def\nn{\nonumber}
\def\lb{\label}
\def\ci{\cite}
\newcommand{\p}{\partial}
\newcommand{\N}{\nabla}

\def\a{\alpha}
\def\b{\beta}
\def\g{\gamma}
\def\d{\delta}
\def\D{\Delta}

\def\e{\epsilon}
\def\o{\omega}

\def\th{\theta}
\def\s{\sigma}

\def\bra{\langle}
\def\ket{\rangle}
\def\l{\left}
\def\r{\right}
\def\f{\frac}
\def\MO{\mathcal {O}}
\def\MR{\mathcal {R}}
\def\MH{\mathcal {H}}
\def\MA{\mathcal {A}}
\def\MM{\mathcal {M}}
\def\MN{\mathcal {N}}

\begin{document}
\title{Black Hole from Entropy Maximization} 
\author{Yuki Yokokura}
\email{yuki.yokokura@riken.jp}
\affiliation{RIKEN iTHEMS, Wako, Saitama 351-0198, Japan}

\begin{abstract}
One quantum characterization of a black hole motivated by (local) holography and thermodynamics is that it maximizes thermodynamic entropy for a given surface area. In the context of quantum gravity, this could be more fundamental than the classical characterization by a horizon. As a step, we explore this possibility by solving the 4D semi-classical Einstein equation with many matter fields. 
For highly-excited spherically-symmetric static configurations, we apply local typicality and estimate the entropy including self-gravity to derive its upper bound. The saturation condition uniquely determines the entropy-maximized configuration: self-gravitating quanta condensate into a radially-uniform dense configuration with no horizon, where the self-gravity and a large quantum pressure induced by the curvatures are balanced and no singularity appears. The interior metric is a self-consistent and non-perturbative solution in Planck's constant. The maximum entropy, given by the volume integral of the entropy density, agrees with the Bekenstein-Hawking formula through self-gravity, deriving the Bousso bound for thermodynamic entropy. Finally, 10 future prospects are discussed, leading to a speculative view that the configuration represents a quantum-gravitational condensate in a semi-classical manner.  
\end{abstract}

\maketitle
\section{Introduction}\label{s:intro}
What is a black hole in quantum theory? The answer is still unknown. Current observational data have not yet shown anything about the interior of a black hole or even confirmed the existence of a horizon \cite{LIGO,EHT,Cardoso}. We have not yet found a theoretical description, fully consistent with quantum theory, that dynamically resolves both the information problem and the singularity, which should be rooted in the quantum nature of gravity. The geometrical characterization of black holes by their horizons was originally based on classical dynamics. However, now that the quantum properties of black holes \cite{Bekenstein, Hawking} have been discovered, it should be natural to consider that black holes are essentially quantum objects consisting of (still unknown) microscopic degrees of freedom, and that there is no a priori reason to follow the classical geometric definition in the context of quantum gravity, where spacetime fluctuates quantum mechanically. 
Therefore, a possible approach is to search for more appropriate and quantum definitions of black holes, to explore their identity, and to study the above problems and quantum gravity. 

One quantum characterization of a black hole is that it maximizes thermodynamic entropy \cite{Dvali3, Oriti1}. This is  not yet fully understood, but it can be motivated by various facts. First, the microscopic origin of thermodynamic entropy is quantum: $S=\log \Omega$, where $\Omega$ is the number of quantum states $\{ |\psi\rangle \}$ consistent with fixed macroscopic quantities. Therefore, the characterization by entropy is valid in a fully quantum context. Second, gravity is universal in that anything with energy attracts each other. In a strong gravity limit, any spherical configuration with mass $\frac{R}{2G}$ will collapse to a black hole with size $R$. This implies that a black hole can be considered as a macroscopic state with maximum entropy according to the second law of thermodynamics. Third, in the case of uncharged spherical symmetry, the Bekenstein-Hawking formula saturates the Bousso bound $S\leq \frac{{\cal A}}{4l_p^2}$ \cite{Bousso1}: a conjecture that the entropy $S$ inside a finite region is bounded by the boundary surface area ${\cal A}$, which has proposed 
holographic principle \cite{
Bousso2} ($l_p\equiv \sqrt{\hbar G}$).\footnote{\label{foot:BH_formula} Note here that the \textit{value} of the Bekenstein-Hawking formula is determined by conserved charges such as ADM energy and does not depend on the presence of a horizon. See also Sec.\ref{s:conclusion}.} 

Then, what macroscopic quantities should we fix for entropy of a self-gravitating system? When calculating entropy of a non-gravitating object microcanonically, one fixes the energy and volume and counts up quantum states consistent with them. In a self-gravitating system, however, energy and volume are related to each other by the Einstein equation, and the two cannot be fixed independently. Rather, motivated by (local) holography \cite{Bousso2, Laurent}, it should be natural to fix the surface area ${\cal A}$ of the boundary of a finite region and consider the configuration that maximizes the entropy \cite{Dvali3,Oriti1}.

Indeed, we can see intuitively that for a spherically-symmetric static system with a fixed surface area ${\cal A}= 4\pi R^2$ $(R\gg l_p)$, self-gravity increases thermodynamic entropy. For highly excited cases, entropy can be estimated as the phase volume, roughly the product of the momentum size and the spatial size \cite{Landau_SM}. The former corresponds to the average kinetic energy, given by temperature. Here, the uniformity (extensivity and intensivity) in the bulk region inside $r=R$ is violated due to the static gravitational field $g_{\mu\nu}(r)$ \cite{Landau_SM,Pad_thermo}, and the local temperature $T_{loc}(r)$ can depend on a position $r$. When Tolman's law holds globally, it is fixed as $T_{loc}(r)=\frac{T_0}{\sqrt{-g_{tt}(r)}}\geq T_0$, where $T_0$ is the temperature of a (nearly) flat region, and $-g_{tt}(r)\leq1$ \cite{Tolman,Landau_SM}. On the other hand, the spatial volume is given by the proper volume $4\pi\int^R_0drr^2\sqrt{g_{rr}(r)}\geq \frac{4\pi R^3}{3}$, where $g_{rr}(r)\geq1$. Thus, the entropy for  ${\cal A}= 4\pi R^2$ is increased by self-gravity $g_{\mu\nu}(r)$\footnote{For example, the entropy of spherical self-gravitating thermal radiation is $\sim R^{\f{3}{2}}$ not $\sim R^3$ \ci{Sorkin}. See Appendix \ref{A:radiation} for a review.} and should be maximized in a strong gravity limit, where a picture of black holes should emerge. 

These suggest a candidate for the quantum definition of black holes: \textit{A black hole is the configuration that maximizes the thermodynamic entropy for a fixed surface area ${\cal A}$.} This is simple and generic in the sense that it depends only on thermodynamic entropy and surface area, and such a configuration can be found in certain quantum-gravity models that have these two notions at some level. 

In this paper, as a first step, we explore this possibility  for spherically-symmetric static configurations with ${\cal A}= 4\pi R^2$ 
in the 4D semi-classical Einstein equation: 
\begin{equation}\label{Einstein}
G_{\mu\nu}=8\pi G \bra \psi|T_{\mu\nu}|\psi \ket,
\end{equation}
the self-consistent equation in a mean-field approximation of quantum gravity where gravity is described by classical metric $g_{\mu\nu}$ and matter by quantum operators  \ci{BD,Kiefer} (here, $R\gg l_p$). 
We then construct the interior metric $g_{\mu\nu}^*$ that maximizes the entropy, and find a candidate picture of quantum black holes. (Here, $X^*$ or $X_*$ denotes a quantity $X$ for the entropy-maximized configuration.) Note that we do \textit{not} assume a priori that a black hole has a horizon or that the maximum entropy is given by the Bekenstein-Hawking formula.

\begin{figure}[h]
\begin{center}
\includegraphics*[scale=0.5]{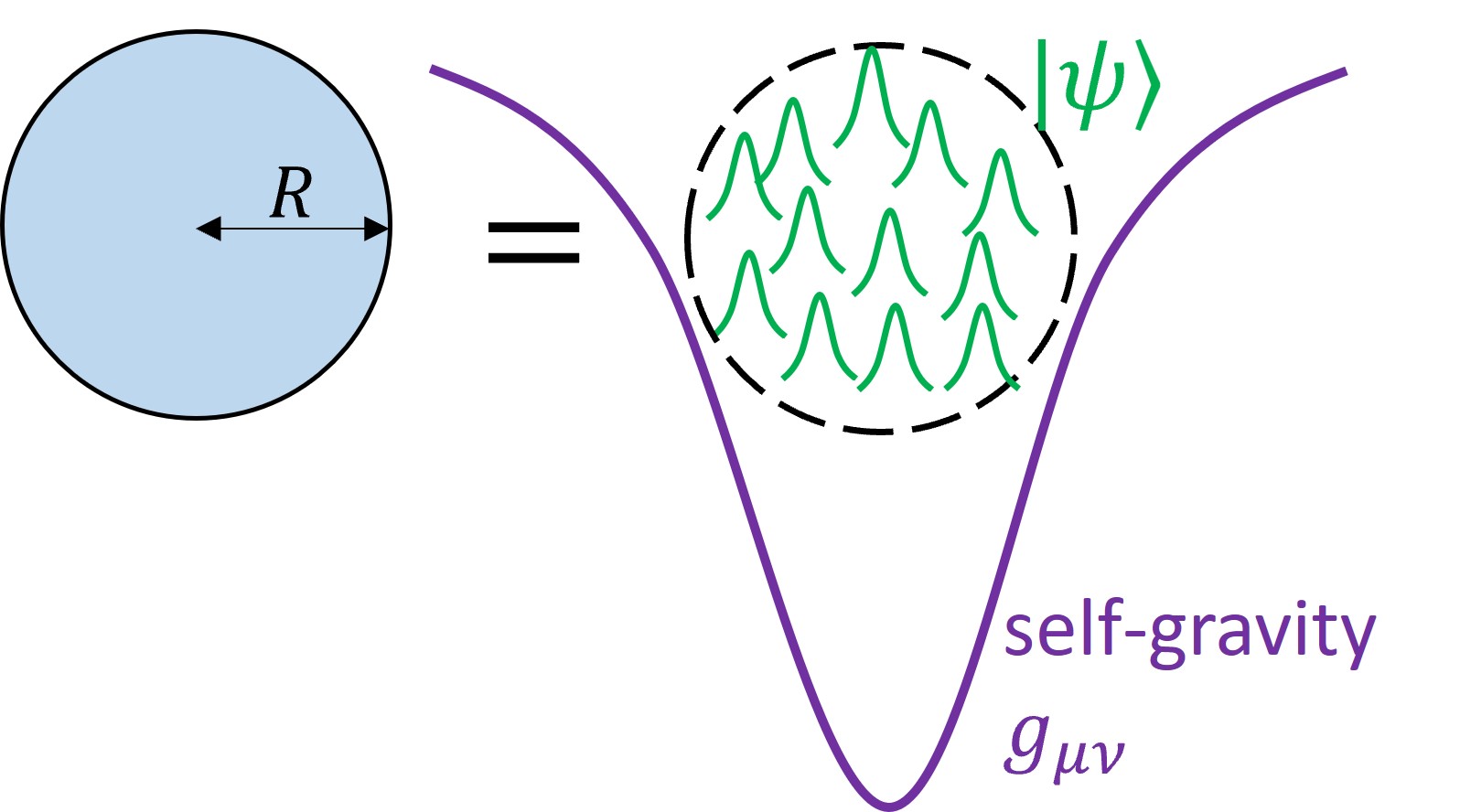}
\caption{A spherical static configuration with size $R$ as a collection of excited quanta in $(g_{\mu\nu},|\psi\ket$) satisfying \eqref{Einstein}.}
\label{f:picture}
\end{center}
\end{figure}

In the framework of \eqref{Einstein}, a spherical static configuration for an excited state $|\psi\ket$ can be considered as a collection of many excited quanta in $|\psi\rangle$ with self-gravity $g_{\mu\nu}$ satisfying \eqref{Einstein} (see Fig.\ref{f:picture}). 
Then, we evaluate the thermodynamic entropy $S$ including the effect of self-gravity $g_{\mu\nu}$ by assuming the phenomenological form \ci{Francesco,Zubarev,FMW,BFM}: 
\begin{equation}\label{S}
S=\int_\Sigma d\Sigma_\mu \tilde s^\mu = \int^R_0 dr \sqrt{g_{rr}(r)}s(r),
\end{equation}
where $\tilde s^\mu$ is the conserved entropy current, $\Sigma$ is chosen as a spacelike hypersurface orthogonal to the timelike Killing vector, and $s(r)$ is the entropy density per proper radial length.\footnote{See \eqref{ss} for the relation of $s$ and $\tilde s^\mu$.} 

Considering the pure-state formulation of \eqref{Einstein} (and a future study of the unitary evolution in the information problem), it should be natural to evaluate the entropy density $s(r)$ by a pure-state method. Here, typicality \cite{Goldstein,Popescu,Sugita,Reimann} states that a subsystem typically behaves as a thermal state in a randomly-chosen state $|\psi\ket$ from a sub-Hilbert space that is consistent with a given macroscopic parameter and has a sufficiently large density of states.
We apply this idea to small subsystems compared to the radius of the curvatures, through the equivalence principle (Sec.\ref{s:entropy}). For a typical and highly excited state $|\psi\rangle$, the energy density $\langle\psi|-T^t{}_t(r)|\psi\rangle$ behaves like a thermodynamic one, 
and the characteristic excitation energy of the typical quanta at $r$ corresponds to the local temperature $T_{loc}(r)$. 
This enables us to estimate the order of the magnitude of $s(r)$. Self-gravity is introduced through the Hamiltonian constraint, which relates $g_{\mu\nu}$ to $s(r)$. As a result, one metric $g_{\mu\nu}$ corresponds to a set of typical states $\{|\psi\rangle\}$ that have the same energy-momentum distribution, and the total entropy \eqref{S} is obtained as a functional of $g_{\mu\nu}$ for a given size $R$:  $S=S[g_{\mu\nu};R)$. It can be seen that the entropy $S$ increases with the excitation energy at each point $r$. 
Here, in order for \eqref{Einstein} to hold, 
we assume that the maximum excitation energy of a quantum is close to but smaller than the Planck energy.  

The upper bound of the entropy then is derived from a semi-classical inequality required by the global static condition: the global existence of the timelike Killing vector (Sec.\ref{s:bound}). Note that static configurations do not have a trapped surface, since the Killing vector would be spacelike inside a trapped surface. Also, the bound leads to the Bekenstein bound \cite{Bek_bound} including self-gravity. 

Solving the saturation condition for the entropy bound under the maximum excitation and using the consistency with local thermodynamics, we find \textit{uniquely} the entropy-maximizing configuration for a fixed surface area ${\cal A}$ (Sec.\ref{s:main}). The interior metric $g_{\mu\nu}^*$ is given by 
\begin{equation}\label{interior0}
    ds^2=-\f{\s\eta^2}{2r^2}e^{-\f{R^2-r^2}{2\s\eta}}dt^2+\f{r^2}{2\s}dr^2+r^2d\Omega^2,
\end{equation}
which is applied for $\sqrt{\s}\lesssim r \leq R$. Two parameters $\s=\MO(n l_p^2)$ and $\eta=\MO(1)$ can be fixed by solving \eqref{Einstein}, where $n$ is a $\MO(1)$ number to be large.\footnote{Here, $\MO(1)$ means $\MO(r^0)$ or $\MO(R^0)$ for $r,R\gg l_p$.} 
We can check that for a theory with many matter fields, 
\eqref{interior0} is a non-perturbative and self-consistent solution of \eqref{Einstein} in $\hbar$, leading to the species bound \ci{Dvali1,Dvali2}. 

\begin{figure}[h]
\begin{center}
\includegraphics*[scale=0.62]{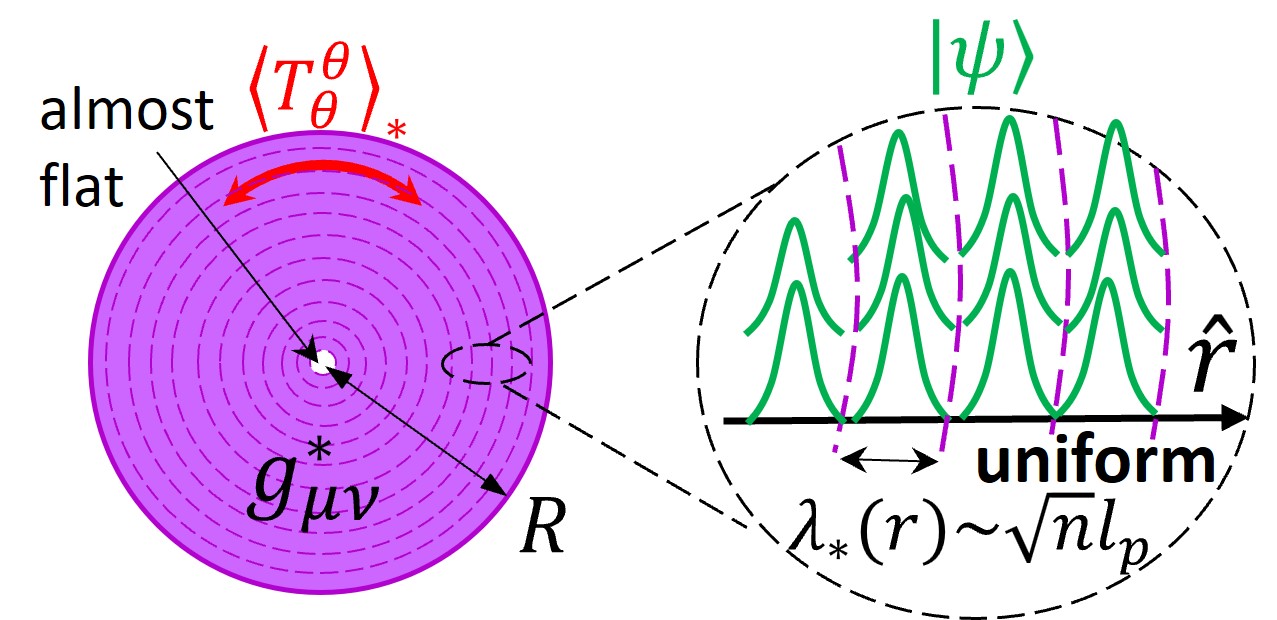}
\caption{Black hole as the semi-classical gravity condensate with the maximum entropy $S_{max}=\frac{\MA}{4l_p^2}$ and the interior metric \eqref{interior0}: a  collection of self-gravitating quanta with wavelength $\lambda_*(r)\sim\sqrt{n}l_p$ distributed uniformly in the proper radial length $\hat r$.}
\label{f:dense}
\end{center}
\end{figure}

Geometrically, \eqref{interior0} is approximately a warped product of $AdS_2$ with radius $\sim \sqrt{n}l_p$ and $S^2$ with radius $r$. Physically, this shows that self-gravitating quanta with near-Planckian excitation energy condensate to form a radially-uniform dense configuration, as in Fig.\ref{f:dense}.
We call it a \textit{semi-classical gravity condensate}. Here, the near-Planckian curvature inside induces quantum fluctuations of various modes, which generate a large tangential pressure $\bra T^\theta{}_\theta \ket_*$ associated with the 4D Weyl anomaly, making the condensate locally anisotropic. This pressure supports the system against the strong self-gravity, exceeds the Buchdahl limit  \ci{Buchdahl}, and self-consistently keeps the curvature finite. As a result, the energy is distributed throughout the interior, and the small central region ($0\leq r \lesssim \sqrt{\s}$) beyond the description by \eqref{Einstein} has only a small energy and can be assumed to be almost flat. Thus, no singularity appears. 

The exterior part $(r\geq R)$ is described approximately by the Schwarzschild metric with ADM mass $\frac{a_0}{2G}(\gg m_p\equiv \sqrt{\hbar/G})$, which is related to the size $R$ as 
\begin{equation}\label{size0}
R=a_{0}+\f{\s\eta^2}{2a_{0}}.
\end{equation}
This is close to but still outside $r=a_0$, and therefore, the configuration has no horizon but looks like a classical black hole from the outside. Indeed, it is almost black due to the exponentially large redshift of \eqref{interior0} \cite{CY}. 

We can also obtain the gravity condensate by considering a formation process. Generically, 
a configuration obtained by a thermodynamically reversible process should 
have the maximum entropy. In Refs.\ci{KMY,KY1,KY4,KY5}, we considered a process in which thermal radiation comes together reversibly due to self-gravity in a heat bath at Hawking temperature, solved the self-consistent time evolution including the backreaction from Hawking-like radiation during the process, and obtained the metric \eqref{interior0}. Therefore, the gravity condensate should be the most typical configuration with the maximum entropy according to the second law of thermodynamics, which is consistent with the above construction based on local typicality. 

Now, to evaluate explicitly $S_{max}\equiv S[g_{\mu\nu}^*;R)$, we apply the Unruh effect (or the local temperature due to the particle creation inside) and thermodynamic relations locally to the interior metric \eqref{interior0} and obtain the entropy density $s_*(r)$ (Sec.\ref{s:BH}). We evaluate \eqref{S} and derive    
\begin{equation}\label{S_bound}
S\leq S_{max}=\f{\MA}{4l_p^2},
\end{equation}
where ${\cal A}\equiv4\pi R^2 = 4\pi a_0^2+\MO(1)$ for \eqref{size0}. 
Therefore, the maximum entropy $S_{max}$ coincides with the Bekenstein-Hawking formula, including the coefficient $1/4$. Here, the self-gravity changes the entropy \eqref{S} from the volume law to the area law \cite{Y1}. We then check the Hawking temperature and the negative specific heat. 

This derives the Bousso bound for thermodynamic entropy (Sec.\ref{s:Bousso}). 
Therefore, \eqref{S_bound} means that the gravity condensate is derived as the unique configuration saturating the Bousso bound in our class of configurations. Furthermore, the interior metric \eqref{interior0} saturates the local sufficient conditions for the Bousso bound proposed in Refs.\ci{FMW,BFM} and should provide a clue to bulk dynamics rooted in the essence of holography.

We finally discuss 10 prospects for this picture of black holes (Sec.\ref{s:conclusion}): role of self-gravity in holography, relation to other gravity-condensate models \cite{Dvali3,Oriti1}, 
description as a thermodynamic phase, 
path-integral evaluation of $S[g_{\mu\nu};R)$ \cite{BY1,BY2,Jacobson_vol}, 
thermodynamic entropy vs entanglement entropy \cite{Minic, Casini, Jacobson_entangle}, recovery of state-dependence in Hawking radiation \cite{KY2}, non-typical configurations, relation to the classical picture of black holes, gravitational field with finite entropy \cite{Jacobson, Pad}, and phenomenology \cite{echo,CY}. 
We then reach a speculative view that the gravity condensate represents a quantum-gravitational condensation in a semi-classical manner.

\section{Estimation of entropy $S[g_{\mu\nu};R)$}\label{s:entropy}
\subsection{Setup}
We start with the setup. Suppose that an excited state $|\psi\rangle$ represents a spherically-symmetric static configuration of size $R$ such that (see Left of Fig.\ref{f:subsystem})
\begin{align}\label{T_psi}
    \langle \psi | T_{\mu\nu}(r) |\psi \rangle = 
    \begin{cases}
    \neq 0 & {\rm for}~~ l_p \ll r \leq R\\
    \approx 0 & {\rm for}~~ R\leq r.
    \end{cases}
\end{align}
Here, we assume that $|\psi\ket$ is excited enough to exceed possible negative energy contributions 
from vacuum fluctuations \ci{BD}, making the total energy density positive ($\langle \psi |- T^t{}_t(r) |\psi \rangle>0$); we exclude a small center region where the semi-classical approximation may break down due to some quantum gravitational effect; and for simplicity, we consider the exterior part approximately vacuum, while a possible large backreaction effect from vacuum fluctuations around the Schwarzschild radius \cite{BD} is taken into account in the interior part. 

Then, we can set the metric by an ansatz:  
\begin{align}\label{metric}
    ds^2 = 
    \begin{cases}
  -\l(1-\f{a(r)}{r}\r)e^{A(r)}dt^2+\l(1-\f{a(r)}{r}\r)^{-1}dr^2+r^2 d\Omega^2 \\ 
  ~~~~~~~~~~~~~~~~~~~~~~~~~~~~~~~~~~~~~~~~~~~~~~~~~{\rm for}~~ l_p \ll r \leq R\\
 -\l(1-\f{a_0}{r}\r)dt^2+\l(1-\f{a_0}{r}\r)^{-1}dr^2+r^2 d\Omega^2\\ 
  ~~~~~~~~~~~~~~~~~~~~~~~~~~~~~~~~~~~~~~~~~~~~~~~~~{\rm for}~~ R\leq r.
    \end{cases}
\end{align}
Here, $\f{a(r)}{2G}$ is the Misner-Sharp mass inside $r$ \ci{Hayward}, and $M_0\equiv \f{a_0}{2G}\approx \f{a(R)}{2G}$ is the ADM energy.\footnote{The ADM energy $M_0$ is not fixed by hand but determined by the mass function $a(r)$ (together with an appropriate junction condition) for a given size $R$.} Eventually, a set $(a(r),A(r), |\psi\ket)$ is determined by solving the semi-classical Einstein equation \eqref{Einstein} self-consistently. (See Appendix \ref{A:EMT} for an example of the self-consistent analysis.) In the following, we suppose such a self-consistent configuration $(a(r),A(r),|\psi \ket)$ satisfying \eqref{Einstein}. 

We now consider this configuration from a microscopic point of view. It consists of a collection of many excited quanta in $|\psi\ket$ as in Fig.\ref{f:picture}. Each excited quantum at $r$ may be in motion, but 
spherically and time averaged, it is stationary with respect to the timelike Killing vector $\p_t$ and has the characteristic excitation energy, $\e(r)(>0)$, measured locally.

\subsection{Local typicality}
\begin{figure}[h]
\begin{center}
\includegraphics*[scale=0.42]{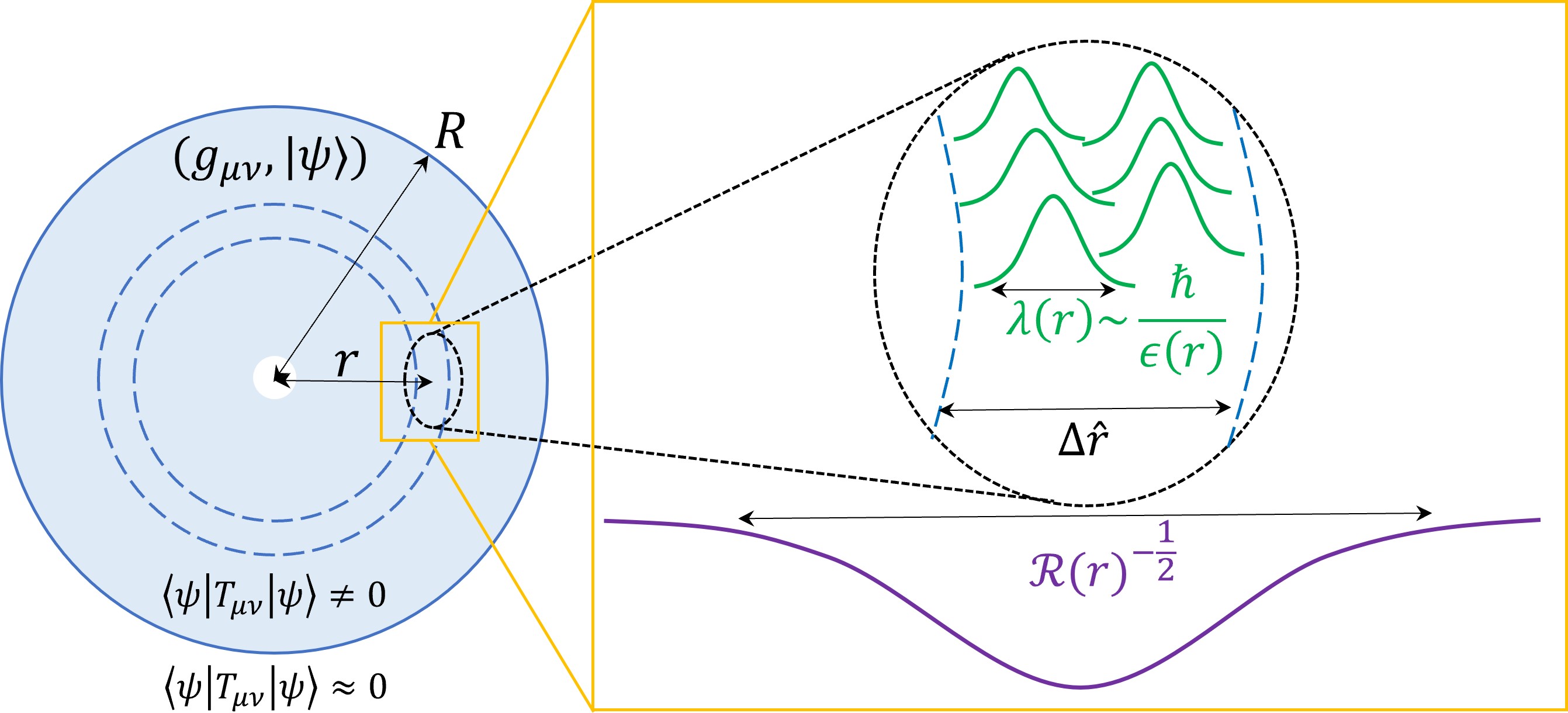}
\caption{Left: A self-consistent configuration $(g_{\mu\nu}, |\psi\ket)$. Right: A spherical subsystem with width $\D \hat r$ where typicality holds.}
\label{f:subsystem}
\end{center}
\end{figure}

\lb{s:setup}
According to the idea mentioned below \eqref{S}, we estimate the entropy density $s(r)$ by using typicality locally.
Let us first focus on a spherical subsystem at $r$ with width $\Delta \hat r\lesssim \MR(r)^{-\f{1}{2}}$ (see Fig.\ref{f:subsystem}). 
Here, $\MR(r)$ denotes the order of the magnitude of the curvatures of the interior metric in \eqref{metric}, and $(\hat t,\hat r)$ is the local coordinate around $r$ with $d \hat t =\sqrt{-g_{tt}(r)}d t$ and $d \hat r =\sqrt{g_{rr}(r)}d r$. Therefore, the bulk of the subsystem can be considered locally flat. If $|\psi\ket$ is sufficiently excited, a quantum around $r$ with $\e(r)$ has a short wavelength such that
\begin{equation}\label{WKB}
\lambda(r)\sim \f{\hbar}{\e(r)}\lesssim \Delta \hat r \lesssim\MR(r)^{-\f{1}{2}}.
\end{equation}

Then, applying typicality \cite{Goldstein,Popescu,Sugita,Reimann} due to the equivalence principle, the spherical subsystem behaves like a local equilibrium in the radial direction such that the local energy $4\pi r^2\bra \psi|T^{\hat t\hat t}(r)|\psi\ket \Delta \hat{r}$ agrees with the thermodynamic one: 
\begin{equation}\label{Ttt_thermal}
   \langle \psi | - T^t{}_t(r) |\psi \rangle \approx \langle - T^t{}_t(r) \rangle_{th}, 
\end{equation}
\footnote{\eqref{Ttt_thermal} is valid only for spherically symmetric averages. We do not consider only configurations that are in local equilibrium in the strict sense, such as those corresponding to local Gibbs states \ci{Zubarev}, nor do we consider only static configurations where Tolman's law holds rigorously \ci{Francesco}. See also Sec.\ref{s:SBH}.} and that $\e(r)$ can be estimated by the local temperature $T_{loc}(r)$:
\begin{equation}\label{e=T}
\e(r)\sim T_{loc}(r),
\end{equation}
since the local temperature governs the local energy scale for a highly excited state.\footnote{We do not consider conserved charges involving chemical potentials.} Here, we use $T^{\hat t\hat t}=-T^{\hat t}{}_{\hat t}=-T^t{}_t$ from $g^{\hat t\hat t}=-1$; $\langle -T^t{}_t(r) \rangle_{th}$ is the thermodynamic energy density given by a function of $T_{loc}(r)$ \cite{Landau_SM}; and $T_{loc}(r)$ is determined self-consistently by \eqref{Einstein} (see Appendix \ref{A:radiation} and Sec.\ref{s:SBH} for examples). 

In general, the order of the magnitude of the entropy in a system with a small volume $\Delta V$ and a high temperature $T$ can be estimated as $\sim \langle - T^t{}_t\rangle_{th} \Delta V /T$. For example, in a spherical ultra-relativistic fluid with volume $4\pi r^2 \Delta \hat{r}$ (including self-gravity), the Stefan-Boltzmann law holds approximately \cite{Landau_SM,fluidbook}, and the entropy is given by $s(r)\Delta \hat{r} =\f{4\bra -T^t{}_t(r)\ket_{th}}{3T_{loc}(r)} 4\pi r^2 \Delta \hat{r}$ (see \eqref{SB_radiation}). Note here that the non-locality of entropy \cite{BFM,BCFM} is considered for the width \eqref{WKB}. 
Therefore, using this, \eqref{Ttt_thermal} and \eqref{e=T}, the order of the magnitude of the entropy in our subsystem can be estimated by $s(r)\Delta \hat r \sim \frac{\langle \psi | - T^t{}_t(r) |\psi \rangle}{\epsilon(r)} 4\pi r^2 \Delta \hat r$, leading to\footnote{One might think that in a pure state, entropy is zero. Microscopically, thermodynamic entropy is given by the number of possible states consistent with fixed macroscopic parameters (most such states are typical), which is state-independent \cite{Landau_SM}. Physically, for example, a cold atomic system in a pure state will develop thermal behavior after a quench process, such that macroscopic quantities such as energy density have the same value as their thermal expectation values. Thus, assuming local typicality, the typical (i.e., thermal) behavior of the energy density can be used to estimate the entropy density in this phenomenological method. (See Sec.\ref{s:conclusion} for microscopic methods.)} 
\begin{equation}\label{s_T}
    s(r)\sim \frac{4\pi r^2 \langle \psi | - T^t{}_t(r) |\psi \rangle}{\epsilon(r)}.
\end{equation}

For the above typicality-based evaluation to be valid, the subsystem must have a sufficiently large density of states, i.e. the entropy of the subsystem must be large: 
\begin{equation}\label{N}
    N(r)\equiv s(r) \Delta \hat r \gg1.  
\end{equation}
Here, using \eqref{s_T} and the total local energy of the subsystem $\Delta E_{loc}\equiv 4\pi r^2 \bra T^{\hat t \hat t}(r)\ket \Delta \hat r$, we have $N(r)\sim \frac{\Delta E_{loc}}{\epsilon(r)}$, which can be considered as the occupation number of excited quanta in the subsystem (see again Fig.\ref{f:subsystem}). Thus, the condition \eqref{N} should hold if we consider a highly excited state and a theory with many degrees of freedom.  As we will see later, this is the case. 

\subsection{Self-gravity}\label{s:S[g]}
We now introduce the effect of self-gravity. 
This is achieved by using the Hamiltonian constraint $\MH=0$ ($G_{tt}=8\pi G \bra \psi| T_{tt}|\psi\ket$) in the interior metric of \eqref{metric}:
\begin{equation}\label{Hami}
\p_ra(r)=8\pi G r^2 \bra\psi|- T^t{}_t(r)|\psi\ket.
\end{equation}
Applying this to \eqref{s_T}, we have 
\begin{equation}\label{s_a}
s(r) \sim \frac{\p_ra(r)}{2G \e(r)},
\end{equation}
which relates geometry and entropy. 
This leads to an interesting expression for the Misner-Sharp mass: 
\begin{equation}\label{a_r}
\f{a(r)}{2G}\sim \int^r_0dr' \e(r')s(r'). 
\end{equation}
From this, the interior metric of \eqref{metric} and \eqref{s_T}, the total entropy \eqref{S} can be estimated as 
\begin{equation}\label{S_typ}
S\sim \int^R_0 dr s(r)\l(1-\f{2G}{r} \int^r_0 dr'\e(r')s(r')\r)^{-\f{1}{2}},
\end{equation}
which provides the entropy $S[g_{\mu\nu}; R)$ for the self-consistent solution to \eqref{Einstein}. Thus, the Hamiltonian constraint \eqref{Hami} relates one metric $g_{\mu\nu}$ and a set of the typical states $\{|\psi\rangle\}$ that have the same thermodynamic energy density $\bra - T^t{}_t(r)\ket_{th}$, and the finite entropy \eqref{S_typ} is obtained.\footnote{\label{foot:thermal}Here are some comments on the subtleties of the above entropy evaluation.We use only the geometrical static condition and local consistencies with thermodynamics, and do not assume a priori a global thermodynamic equilibrium. This is motivated by two facts. More physically, even for static configurations, one should consider their formation processes. Generically, mechanical and global/local thermodynamic equilibria are different due to differences in relaxation timescales \ci{Landau_SM}. In particular, we are now considering self-gravity. Therefore, it is necessary to consider time-delay effects in the formation process and discuss in which equilibrium state the configuration is for the timescale under consideration. The other is that, a notion of global thermodynamic equilibrium in self-gravitating systems, leading often to thermodynamical instability from negative specific heat \ci{Antonov,Lynden}, is still controversial \ci{Landau_SM, Pad_thermo}, and in particular, one consistent with \eqref{Einstein} is not known. Furthermore, we do not assume a condition of isotropic fluid \cite{Green,Xia} (like thermal radiation \cite{Sorkin}) because it is not clear a priori whether such fluid maximizes entropy including self-gravity for a fixed surface area. Thus, we have estimated $S[g_{\mu\nu};R)$, \eqref{S_typ}, by using only the geometric static condition and the local consistency with thermodynamics. For the case of $S_{max}$, we will check the validity of this treatment by constructing the self-consistent solution $(g_{\mu\nu}^*,|\psi\ket_*)$ (Sec.\ref{s:g*}). In Sec.\ref{s:conclusion} we will discuss a global equilibrium in quantum gravity briefly. In Appendix \ref{A:Tolman} we will argue that Tolman's law does not lead to maximum entropy.}  

The estimation \eqref{S_typ} shows that for a given $s(r)$, the largest $\e(r)$ at each $r$ leads to the largest $S$. More precisely, according to the second law within each spherical subsystem, the entropy density $s(r)$ should be an increasing function of the local temperature $T_{loc}(r)$ \ci{Landau_SM}, and from \eqref{e=T}, 
the maximum local temperature at each $r$ provides the maximum entropy $S_{max}$ for a given size $R$. This is consistent with ordinary thermodynamics without self-gravity \ci{Landau_SM}, but it is a non-trivial result because self-gravity is included here. 

Then, what is the maximum excitation energy? 
In order for the semi-classical description to be valid, the characteristic excitation energy $\e(r)$ (or the local temperature $T_{loc}(r)$ from \eqref{e=T}) must satisfy \ci{Pad_Lim,Caianiello,Brandt}
\begin{equation}\label{e_max}
\e(r)\leq  \e_{max} \sim \f{m_p}{\sqrt{n}}
\end{equation}
with $n=\MO(1)\gg1$, a large number to be determined. 

\section{Upper bound}\label{s:bound}
\begin{figure}[h]
\begin{center}
\includegraphics*[scale=0.42]{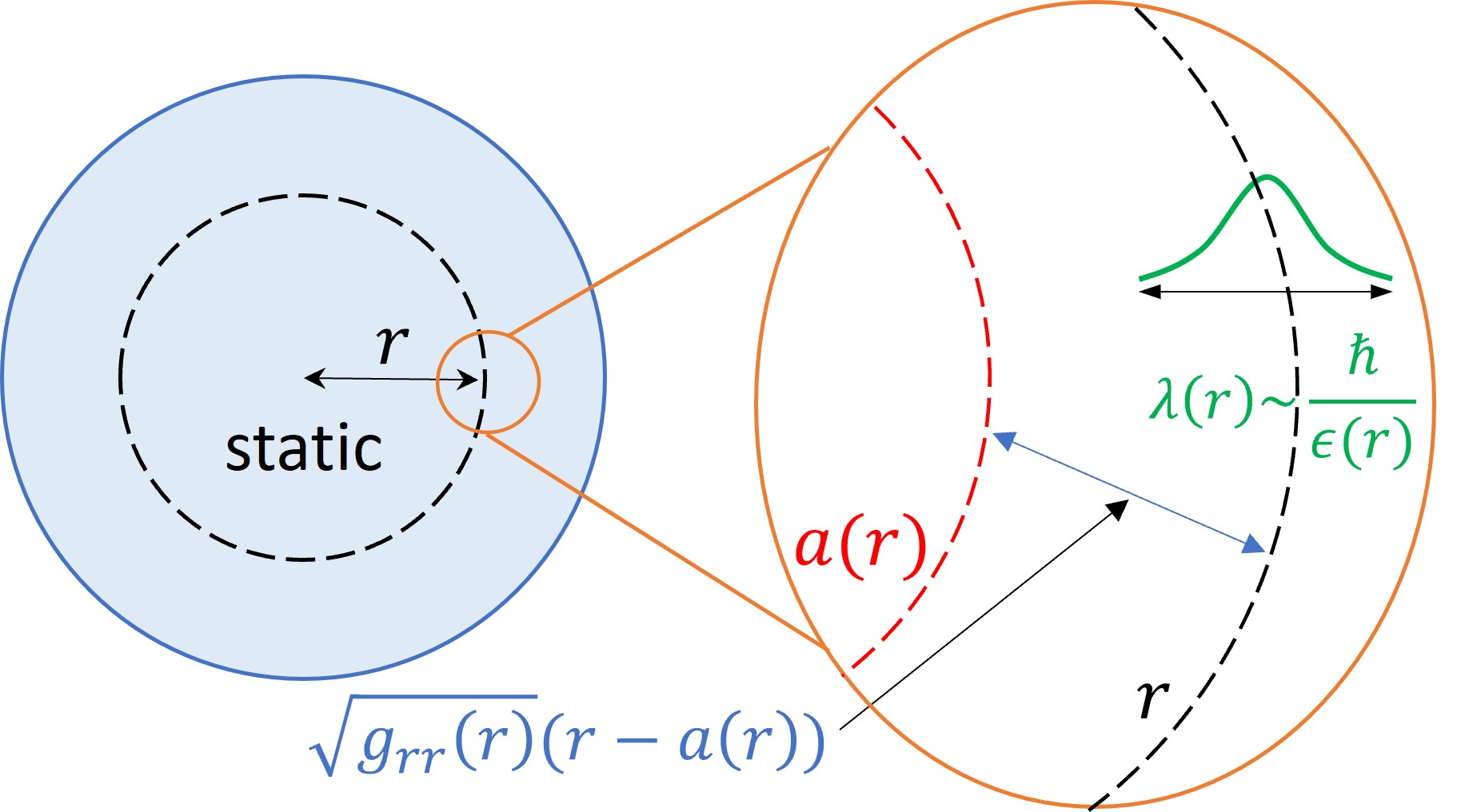}
\caption{Meaning of the semi-classical static condition \eqref{no_trap}.}
\label{f:static}
\end{center}
\end{figure}

We derive the upper bound for $S[g_{\mu\nu};R)$. 
A static spacetime has a timelike Killing vector globally, 
indicating that there is no trapped surface \ci{Bousso2,No_trap}. 
This condition can be expressed at a semi-classical level as 
\begin{equation}\label{no_trap}
\lambda(r)\lesssim \sqrt{g_{rr}(r)}(r-a(r))
\end{equation}
in the interior metric of \eqref{metric} \cite{Sorkin}. See Fig.\ref{f:static}. 
This means that, considering the non-locality of quanta, a quantum that constitutes a part at $r$ of a static configuration must be located at least its wavelength away from the Schwarzschild radius $a(r)$ of the energy inside it; otherwise, the trapped region would appear, the quantum would lie inside it and fall toward the center, and the configuration would not be static. (The validity of \eqref{no_trap} will be discussed in Sec.\ref{s:formation}.) 

Using \eqref{s_T} and $\lambda\sim\f{\hbar}{\e}$, 
we can rewrite \eqref{no_trap} as 
\begin{equation}\lb{no_trap_Bek}
s(r)\lesssim \f{1}{\hbar}\sqrt{g_{rr}(r)}(r-a(r)) 4\pi r^2 \bra \psi| -T^t{}_t(r)|\psi \ket,
\end{equation}
whose meaning will be discussed in Sec.\ref{A:Bousso1}.
From this, \eqref{S} and \eqref{metric}, we can calculate 
\begin{align}\label{S_B0}
S &\lesssim \f{4\pi}{\hbar} \int^R_0 dr r^2 g_{rr}(r)(r-a(r)) \bra\psi| -T^t{}_t(r)|\psi\ket \nn \\
 &= \f{4\pi}{\hbar} \int^R_0 dr r^3 \bra\psi| -T^t{}_t(r)|\psi\ket.
\end{align}
We apply \eqref{Hami} and obtain the upper bound: 
\begin{equation}\label{S_B}
S \lesssim \f{1}{l_p^2}\int^R_0 dr r \p_r a(r),
\end{equation}
which holds universally for any configuration in our class. 

Here, we can derive from \eqref{S_B0} the Bekenstein bound \ci{Bek_bound} including self-gravity: 
\begin{equation}\label{S_BB}
S \lesssim \f{R M_0}{\hbar},
\end{equation}
where we have used
$4\pi\int^R_0 dr r^3 \bra -T^t{}_t(r)\ket\leq 4\pi R \int^R_0 dr r^2 \bra -T^t{}_t(r)\ket=R \f{a(R)}{2G}\approx R \f{a_0}{2G}\equiv RM_0$ 
from \eqref{Hami}.\footnote{In Ref.\ci{Sorkin}, \eqref{S_B0} and \eqref{S_BB} were obtained only for thermal radiation, 
while our derivation can be applied to more general cases.} 
Note that $G$ does not appear here, 
but this is a result of the dynamics of gravity because we need $\MH=0$ \eqref{Hami} to obtain the ADM energy $M_0$. 

\section{Entropy-maximized configuration}\label{s:main}
We take a strong-gravity limit in a consistent way and find the entropy-maximized configuration $g_{\mu\nu}^*$ that saturates the bound \eqref{S_B}, to obtain a quantum picture of black holes. Here, \eqref{S_B} is just an order estimate, but remarkably, the saturation condition and the consistency with our arguments so far can determine the functional form of $g_{\mu\nu}^*(r)$ uniquely, except for two constant parameters. They can be fixed by solving \eqref{Einstein} self-consistently.

\subsection{Saturating energy distribution $a_*(r)$}\label{s:a*}
Let us first find the energy distribution $a_*(r)$ that saturates the inequality \eqref{S_B}. 
Squaring the saturation condition for \eqref{no_trap} 
and using \eqref{metric} and $\lambda(r)\sim \f{\hbar}{\e(r)}$, 
we have 
\begin{equation}\label{eq_a*}
\hbar^2 \e_*(r)^{-2}\sim r(r-a_*(r)).
\end{equation}
The estimation \eqref{S_typ} and bound \eqref{e_max} mean that 
the maximum entropy can be obtained by setting $\e_*(r)=\e_{max}$ at each $r$. 
Therefore, \eqref{eq_a*} becomes $n l_p^2\sim r(r-a_*(r))$, leading to 
\begin{equation}\label{a_*}
a_*(r)=r-\f{2\s}{r}
\end{equation}
with $\s=f n l_p^2$ ($f$: a dimensionless positive constant of $\MO(1)$).
Thus, the maximum entropy $S_{max}\equiv S[g_{\mu\nu}^*,R)$ is estimated from \eqref{S_B} and \eqref{a_*} as 
\begin{equation}\label{S_*}
S_{max}\sim \f{1}{l_p^2}\int^R_0dr r \p_r a_*(r) \sim \f{R^2}{l_p^2}\approx \f{a_0^2}{l_p^2}, 
\end{equation}
where $a_0\approx a_*(R)\approx R$. 
(In Sec.\ref{s:surface}, the precise relation of $a_0$ and $R$ will be fixed.) Note that the saturation of both \eqref{e_max} and \eqref{no_trap} corresponds to the strong-gravity limit mentioned in Sec.\ref{s:intro}. 

$a_*(r)$ represents a radially uniform configuration in that the entropy density is constant: 
\begin{equation}\label{s*}
    s_*(r)\sim \frac{\sqrt{n}}{l_p},
\end{equation}
where we have applied $\e(r)=\e_{max}$ and \eqref{a_*} to \eqref{s_a}. This means that $\MO(\sqrt{n})$ bit of information is packed per the Planck length constantly \ci{KY2,Y1}. Thus, a necessary condition for the entropy maximization is the radial uniformity. Conversely, as a result of the radial uniformity, \eqref{a_*} can be uniquely obtained as a solution satisfying the entropy maximization condition \eqref{eq_a*}, without using $\e(r)=\e_{max}$ (see Appendix \ref{A:N(r)}). Therefore, from these two arguments, a necessary and sufficient condition for the entropy maximization is the radial uniformity. This is a non-trivial result for a self-gravitating system.

We here examine in which region the energy distribution \eqref{a_*} is valid. As we will see in \eqref{Dr*}, the width $\Delta \hat r _*$ of the subsystem is almost the same as $\lambda_*(r)\sim \frac{\hbar}{\epsilon_{max}}\sim \sqrt{n}l_p$. From this, \eqref{s*} and the assumption $n\gg1$, then the number \eqref{N} of excited quanta in each subsystem is large indeed: 
\begin{equation}\label{N*}
    N^*(r)\sim n \gg 1. 
\end{equation}
On the other hand, using \eqref{Hami} and \eqref{a_*}, we have the energy density $\bra-T^t{}_t\ket_*\approx\frac{1}{8\pi G r^2}$ and the local energy in each subsystem, $\Delta E_{loc}^*(r)\equiv 4\pi r^2 \bra-T^t{}_t\ket_* \Delta \hat r_* \sim \sqrt{n} m_p$. 
Therefore, we must have at least $\f{a_*(r)}{2G}\gtrsim \sqrt{n}m_p$, 
meaning that \eqref{a_*} holds only in 
\begin{equation}\label{a*_range}
\sqrt{n}l_p \lesssim r \leq R.
\end{equation}
For $n\gg1$, this is consistent with the assumption we have made below \eqref{T_psi}. 

\subsection{Determination of the interior metric $g_{\mu\nu}^*$}\label{s:g*}
We determine the metric $g_{\mu\nu}^*$. First, \eqref{a_*} fixes $g_{rr}^*(r)=\f{r^2}{2\s}$ in the interior metric of \eqref{metric}. Next, from \eqref{a*_range} and the condition $n \gg 1$, we can focus on the asymptotic form $A(r)=C r^k$ for $r\gg l_p$. Here, it should be natural to assume that there is no length scale in $g_{\mu\nu}^*$ except for $l_p$, since the saturating configuration consists of excited quanta with $\e_{max} \sim \f{m_p}{\sqrt{n}}$. Then, $C$ is a constant of $\MO(1)$ which can depend on $l_p$ but not on $a_0$. 
Indeed, this will lead to a self-consistent solution of \eqref{Einstein}. 

To find a physical value of $k$, we calculate 
\begin{align}
\lb{Gtt}
-G^t{}_t &=\frac{1}{r^2}+\frac{2 \sigma }{r^4},\\
\lb{Grr}
G^r{}_r &=-\frac{1}{r^2}-\frac{2 \sigma }{r^4}+ 2 C k \sigma  r^{k-4},\\
\lb{G33}
G^\theta{}_\theta &= \frac{2 \sigma }{r^4}+C (k-3) k \sigma  r^{k-4}+\frac{1}{2} C^2 k^2 \sigma  r^{2 k-4},
\end{align}
and use two self-consistencies:  thermodynamics and semi-classicality.  
Because we have supposed quanta consistent with local thermodynamics, the pressures must be positive \ci{Landau_SM}.
This and \eqref{Grr} require $C>0$ and $k-4\geq-2$, that is, $k\geq2$. 
Also, we are now considering 
the semi-classical condition \eqref{e_max}, 
which means that the curvatures $\MR(r)$ must be at most of $\MO(r^0)$ for $r\gg l_p$. 
This restricts the highest term, the third one of \eqref{G33}, such that $2k-4\leq0$, that is, $k\leq2$. 
Therefore, we can only have $k=2$. 
From dimensional analysis, we can set $C=\f{1}{2\s\eta}$ with a $\MO(1)$ dimensionless  constant $\eta>0$. 

Thus, we have reached \textit{uniquely} the interior metric $g_{\mu\nu}^*$ with the maximum entropy in the class of spherically-symmetric static spacetime: 
\begin{equation}\label{interior}
ds^2=-\f{2\s}{r^2}e^{\f{r^2}{2\s\eta}+A_0}dt^2+\f{r^2}{2\s}dr^2+r^2 d\Omega^2,
\end{equation}
which is valid for the range \eqref{a*_range}. $A_0$ is a constant for connection to the exterior metric (see \eqref{interior2}).

Note that \eqref{interior} can be obtained in many ways: adiabatic formation in a heat bath of Hawking temperature  \ci{KMY,KY1,KY4,KY5}, consistency with the Bekenstein-Hawking formula \ci{Y1}, and typical configuration in a solution space of \eqref{Einstein} \ci{HKLY}. 
In this sense, \eqref{interior} is robust, and the present argument provides 
a derivation that characterizes it as the entropy-maximizing configuration.

We now check that the metric \eqref{interior} satisfies \eqref{Einstein} self-consistently. We here give an outline of the proof. (See Appendix \ref{A:EMT} for a short review and Ref.\ci{KY4} for the details.) We first consider, say, $n_s(\gg1)$ scalar fields in the background metric \eqref{interior} and solve the matter field equations with a perturbative technique by employing the fact that the metric is a warped product of $AdS_2$ with radius $L\equiv \sqrt{2\sigma\eta^2}$ and $S^2$ with radius $r$ (see the Ricci scalar in \eqref{RRR}).\footnote{\label{foot:AdS}More precisely, we have $R_{RicciScalar}=-\f{2}{L^2}+\f{2}{r^2}$, and we can check that the metric \eqref{interior} is equivalent locally to $ds^2=\f{L^2}{z^2}(-dt^2+dz^2)+r(z)^2d\Omega^2$ \ci{KY4}.} We then use dimensional regularization to evaluate the renormalized energy-momentum tensor $\bra\psi|T_{\mu\nu}|\psi\ket_*$ for a typical state $|\psi\ket_*$ in which only s-waves are excited with $\sim \e_{max}$ and the other modes are in the ground state $|0\ket$ of \eqref{interior}.\footnote{Note that $|\psi\ket_*$ is not the local Gibbs state \ci{Zubarev} with $T_{loc}\sim\e_{max}$.} 
We finally compare both sides of \eqref{Einstein} to find the self-consistent values: 
\begin{equation}\label{sigma}
\s=\f{n_s l_p^2 }{120 \pi \eta^2},~~1 \leq \eta<2.
\end{equation}
We can check that \eqref{sigma} is consistent with 4D Weyl anomaly \ci{BD} (see Appendix \ref{A:EMT}). 
Thus, we conclude that \eqref{interior} with \eqref{sigma} is the non-perturbative solution of \eqref{Einstein} with $n_s$ scalar fields in the sense that the limit $\hbar\to0$ cannot be taken in \eqref{interior} and \eqref{RRR}. 

We here discuss two points about the solution.
\subsubsection{Meaning of $n$ and species bound}
We discuss the meaning of $n$. 
Comparing $\s=f n l_p^2$ to \eqref{sigma}, we have $n=n_s$, the number of scalar fields. In the case where matter fields are conformal, we can determine the self-consistent value of $\sigma$, which is different from \eqref{sigma} and shows $n=c_W$ \ci{KY1,KY3,KY5}.  
Here, $c_W$ is the coefficient of the square of Weyl tensors in the 4D Weyl anomaly \ci{BD}. 
Therefore, $n$ represents the number of the degrees of freedom in the theory that can contribute to entropy (see \eqref{s*} and \eqref{s2}). If we consider a theory with many species of fields, the condition $n\gg1$ is satisfied. 
On the other hand, $n$ has been introduced in \eqref{e_max} as a parameter characterizing the maximum energy $\e_{max}$ 
for which a semi-classical description is valid. 
Thus, \eqref{e_max} agrees with the species bound \ci{Dvali1}. 
This is the result of solving \eqref{Einstein} non-perturbatively, which is another derivation of the species bound. Note that the species bound also appears from the validity of the Bekenstein bound \ci{Dvali2}, while we first derive the upper bound \eqref{S_B} and then reach the species bound. This may imply that there is an intrinsic relationship between the species bound and the entropy bound. 

\subsubsection{No singularity}\label{s:No_sing}
We can expect that there is no singularity. 
First, note that the interior metric \eqref{interior} is valid only in the range \eqref{a*_range}. In a large $n$, we have the leading terms of the curvatures for $r\gg l_p$:
\begin{equation}\label{RRR}
R_{\rm RicciScalar}=-\f{2}{L^2},~~R_{\mu\nu}R^{\mu\nu}=\f{2}{L^4},~~R_{\mu\nu\a\b}R^{\mu\nu\a\b}=\f{4}{L^4},
\end{equation}
where $L\equiv \sqrt{2\s\eta^2} \sim \sqrt{n}l_p \gg l_p$. 
These are close to but still smaller than the Planck scale: $\MR_*(r)\sim \f{1}{n l_p^2}$. On the other hand, we have 
$\lambda_*(r)\sim \f{\hbar}{\e_{max}}\sim \sqrt{n} l_p$. Therefore, 
the width of the subsystems is minimum:  
\begin{equation} \label{Dr*}
\Delta \hat r _* \sim \lambda_*(r) \sim \MR_*(r)^{-\f{1}{2}} \sim \sqrt{n}l_p,
\end{equation}
which satisfies the condition \eqref{WKB}, albeit barely. 

We next examine the small center region $0\leq r \lesssim \sqrt{n}l_p$, which cannot be described by the metric \eqref{interior}. 
The energy inside is estimated through \eqref{a_*} as $\f{a_*(\sqrt{n}l_p)}{2G}\sim\sqrt{n}m_p$, which is much smaller than $\f{a_0}{2G}$, one expected in a classical case. Therefore, the center region cannot have a singularity appearing in the classical cases, and rather it should correspond to a small excitation of quantum-gravitational degrees of freedom like string (which can be confirmed only by a future development). In this paper, we assume that the center part is almost flat.\footnote{The metric \eqref{interior} becomes flat around $r\approx \sqrt{2\s}$: $ds^2|_{r\approx \sqrt{2\s}}\approx -e^{\f{1}{\eta}+A_0}dt^2+dr^2+r^2 d\Omega^2 \nn$,
which is a flat metric by redefining time coordinate $t$. Also, we can consider the time evolution from formation to evaporation and see that the center region is kept flat, except for the final stage of the evaporation \ci{KY4}.} 

Thus, the non-perturbative solution \eqref{interior} with \eqref{sigma} has no singularity. We will see in Sec.\ref{s:pressure} that a quantum repulsive force is generated inside and plays a key role in the resolution of the singularity. 

\subsection{Semi-classical gravity condensate}\label{s:dense}
We study the configuration of \eqref{interior}. 
(See Ref.\ci{KY2} for aspects not discussed here.)  
First, it has through \eqref{Einstein}
\begin{align}\label{EMT}
\bra-T^t{}_t(r)\ket_*=\frac{1}{8\pi Gr^2},&~~\bra T^r{}_r(r)\ket_* =\f{2-\eta}{\eta}\bra -T^t{}_t(r)\ket_*,\nn\\
\bra T^\theta{}_\theta (r)\ket_*&=\f{1}{16\pi G\eta^2\s},
\end{align}
as the leading terms for $r\gg l_p$. 
Note again that these are applied only to the region \eqref{a*_range}.
First, the energy density $\bra-T^t{}_t\ket_*$ is positive as a result of the contributions both from the excited quanta and vacuum fluctuations \ci{KY4}. 
Second, we have $\bra T^r{}_r\ket_*>0$ for $\eta$ satisfying \eqref{sigma}, and $\eta$ can be considered as the parameter in the equation of state. 
Third, $\bra T^\theta{}_\theta \ket_*=\MO\l(\f{1}{G nl_p^2}\r)$ is close to the Planck scale, large enough to violate the dominant energy condition; it is locally anisotropic (\textit{not} fluid) enough to exceed the Buchdahl limit  \ci{Buchdahl}; 
and it supports the system against the strong self-gravity.\footnote{$\N_\a \bra T^\a{}_r(r)\ket=0$ in the interior metric of \eqref{metric} gives the anisotropic TOV equation: $0=\p_r \bra T^r{}_r \ket + \p_r \log \sqrt{-g_{tt}}(\bra -T^t{}_t \ket +\bra T^r{}_r \ket)
+\f{2}{r} (\bra T^r{}_r \ket -\bra T^\theta{}_\theta \ket )=0$. 
Using \eqref{interior} and \eqref{EMT}, we have as the leading ones for $r\gg l_p$ 
\begin{align*}
\N_\a \bra T^\a{}_r\ket&\approx\f{r}{2\s \eta}\l(1+\f{2-\eta}{\eta}\r) \bra- T^t{}_t \ket
+\f{2}{r} (-\bra T^\theta{}_\theta \ket )\\
 &= \f{r}{2\s \eta} \f{2}{\eta} \f{1}{8\pi G r^2} - \f{2}{r} \f{1}{16\pi G \s\eta^2}=0.  
\end{align*}
This shows that the tangential pressure supports the configuration against the strong self-gravity. } 

Noting the radial uniformity (discussed in Sec.\ref{s:a*}), we thus reach the picture shown in Fig.\ref{f:dense}. The metric \eqref{interior} represents a self-gravitating condensate consisting of the excited quanta with $\epsilon_{max}$, distributed uniformly in $\hat r$ direction, and the vacuum fluctuations (\textit{semi-classical gravity condensate}). It is dense in the sense that it has the large pressure and curvatures. The exterior geometry ($r\geq R$) is approximated by the Schwarzschild metric in \eqref{metric}, where the curvature is small. We can check here that the curvatures jump at the surface (located at \eqref{size}) in a mild manner to keep as much the interior uniformity as possible, 
consistent with Israel's junction condition \ci{Poisson} (see Sec.7 in Ref.\ci{Y1} for the details). 

Let us now find a rough form of $A_0$ and consider what it means. First, we note from \eqref{a_*} that $a_0 \approx a_*(R) =R-\f{2\s}{R}$, indicating that the surface exists at $r=R=a_0 + \MO\l(\f{n l_p^2}{a_0}\r)$ (see \eqref{size} for the precise one). For the interior metric \eqref{interior} to connect at $r=R$ to the exterior Schwarzschild metric in \eqref{metric}, 
the induced metric on $r=R$ must be continuous \ci{Poisson}, which requires $-g_{tt}(R)=1-\f{a_0}{R}\approx \frac{\MO(nl_p^2)}{R^2}=\f{2\s}{R^2}e^{\f{R^2}{2\s\eta}+A_0}$. Then, we obtain 
\begin{equation}\lb{gtt}
g_{tt}^*(r)=-\f{2K\s}{r^2}e^{-\f{R^2-r^2}{2\s\eta}}, 
\end{equation}
where $K$ is a $\MO(1)$ number to be fixed. This is an exponentially large redshift \cite{KMY}. To see this effect, setting $r=R-\D r$ ($\D r \ll R$), \eqref{gtt} becomes $-g_{tt}^*(r)\sim e^{-\f{R \D r}{\s\eta}}$. This indicates that due to the strong redshift, a deep region with $\D r \gg \MO\l(\f{n l_p^2}{a_0}\r)$ is almost frozen in the outside time $t$. As a result, 
the information carried by the excited quanta is kept inside for an exponentially long time \ci{KY2}, and the gravity condensate looks almost black. (We will discuss its phenomenological effect in Sec.\ref{s:conclusion}.)

\subsubsection{Origin of \eqref{no_trap} and formation process}\label{s:formation}
We have started from the semi-classical static condition \eqref{no_trap} and reached uniquely the entropy-maximizing metric \eqref{interior}. We here discuss the origin of \eqref{no_trap}. 

First, as discussed around Fig.\ref{f:static}, in order for the whole region to be static, there cannot exist a trapped region, and  the condition \eqref{no_trap} should hold as a physical condition that takes quantum effects into account. In fact, we can check explicitly that this is the case for specific static spherically symmetric metrics such as self-gravitating thermal radiation (see Appendix \ref{A:radiation}).

Second, a possible origin of \eqref{no_trap} is a local sufficient condition for the Bousso bound, Eq.(1.9) in Ref.\ci{FMW} (given by \eqref{FMW3} in our paper). We note that the entropy-density inequality \eqref{no_trap_Bek} is equivalent to the condition \eqref{no_trap} for highly-excited configurations in which the relation \eqref{s_T} holds. In Sec.\ref{A:Bousso1}, we will show that the inequality \eqref{no_trap_Bek} is a spherical static version of \eqref{FMW3}. Therefore, the static condition \eqref{no_trap} is valid for highly-excited spherical static configurations where \eqref{FMW3} holds. Note that  \eqref{FMW3} is a hypothesis motivated by a phenomenological argument \ci{FMW}, and that it is not clear yet which is more fundamental, \eqref{no_trap} or \eqref{FMW3}. 

\begin{figure}[h]
\begin{center}
\includegraphics*[scale=0.55]{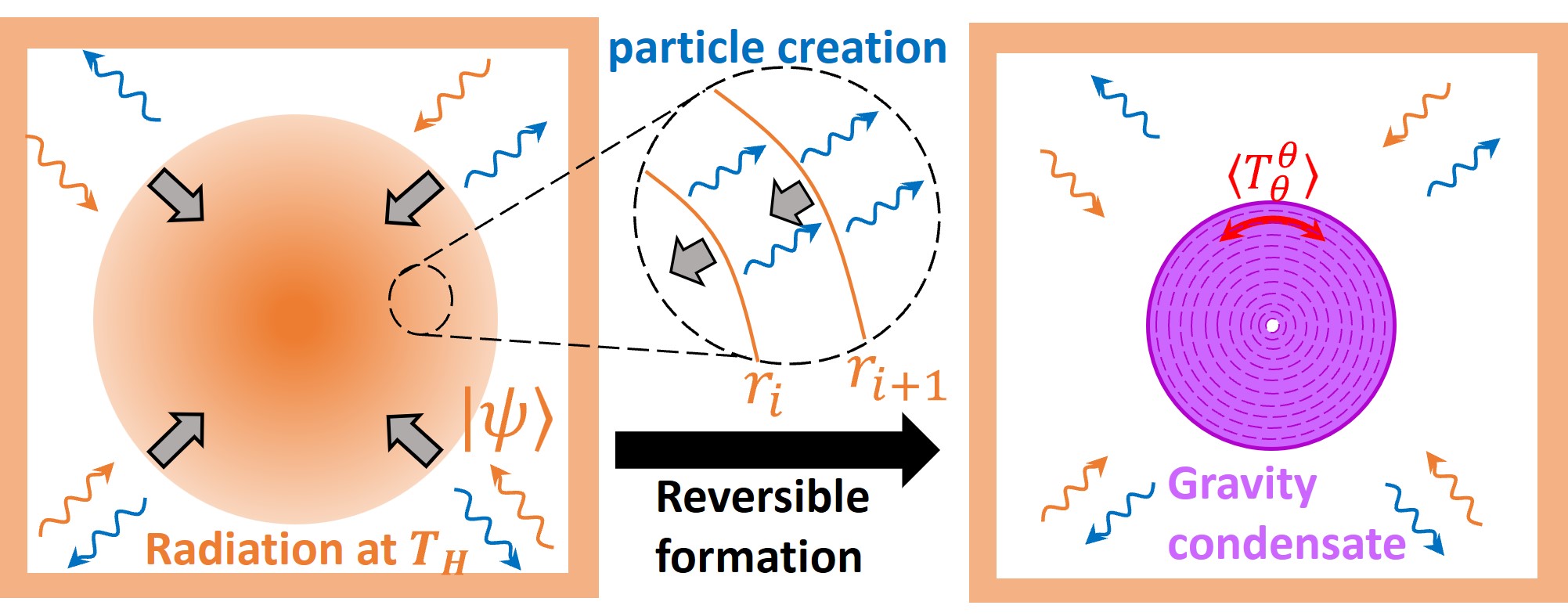}
\caption{Reversible formation of the semi-classical gravity condensate in a heat bath of Hawking temperature.}
\label{f:pressure}
\end{center}
\end{figure}

Third, in order to explore a dynamical origin of \eqref{no_trap}, let us consider the formation process of the gravity condensate and investigate how the condition \eqref{no_trap} is saturated as a result of the time evolution. 
We first note that the gravity condensate has, by construction, the maximum entropy and should be obtained by a reversible process in thermodynamics 
(since, in general, entropy is kept at the maximum for the external parameters given at each step of such processes). 
Suppose, then, that a collection of many spherical radiations at Hawking temperature comes together slowly due to self-gravity. See Fig.\ref{f:pressure}. 
This corresponds to a reversible formation process in a heat bath and can be realized operationally by slowly changing the temperature of the heat bath to match the Hawking temperature of the energy at each stage \ci{KY1}. 

We can use \eqref{Einstein} and make a self-consistent model that describes this process \ci{KMY} 
(see Appendix in Ref.\cite{KY5} for a short review). We first note that in this formation process, the metric changes in time, and that generically, particle creation occurs in a time-dependent spacetime \ci{BD,Barcelo}. As in Fig.\ref{f:pressure}, we model the radiation as a collection of concentric spherical null shells with small energy $\frac{\Delta a_i}{2G}$ and describe the region between the $i$-th and $(i+1)$-th shells by the Vaidya metric with energy $\frac{a_i}{2G}$, corresponding to the Bondi energy inside the $i$-th shell with radius $r_i$ (here, $\Delta a_i\equiv a_i- a_{i-1}>0$ and $r_i < r_{i+1}$). We can solve the WKB-approximated time evolution of s-waves of $n_s$ scalar fields that come in the Minkowski vacuum state and propagate through the metric, to build a formula for the energy flux of the particles created around each shell ($n_s=\MO(1)\gg1$). Then, by solving \eqref{Einstein} with the flux formula self-consistently, we can analyze the time evolution of both the shrinking shells and the geometry,\footnote{\label{foot:backreaction}It is often believed that the backreaction from evaporation is negligible when considering the formation process of a black hole with mass $\frac{a_i}{2G}$ since the timescale of the evaporation $\Delta t \sim \frac{a_i^3}{l_p^2}$ is much longer than the timescale of the collapse $\Delta \tau \sim a_i$. However, this naive expectation is wrong because $t$ and $\tau$ are the time coordinate at infinity and that of the comoving observer along the collapsing matter, respectively, and it does not make sense to compare such two timescales. In fact, one can examine the time evolution of the evaporation and collapse in a common time coordinate, and check that the both effects compete near $r=a_i$ and the evolution \eqref{r_a_i} occurs \cite{KMY,KY4}.} to see that around each shell, Hawking-like radiation occurs 
with temperature $\frac{\hbar}{4\pi a_i}$ and intensity $\sigma_s \sim n_s l_p^2$, and that each shell approaches just outside the shrinking Schwarzschild radius $a_i$ due to the evaporation,  
\begin{equation}\label{r_a_i}
    r_i \to a_i + \frac{2\sigma_s}{a_i}, 
\end{equation}
without forming a trapped region. 
Taking a continuum limit $\Delta a_i \to 0$, we can construct the interior metric \eqref{interior} with $\sigma=\sigma_s$ and $\eta=1$, indicating that the gravity condensate is formed in this evolution. 

From the evolution \eqref{r_a_i}, we can see a dynamical saturation of \eqref{no_trap}. When the size of the gravity condensate is $\sim a_i$, the temperature of the bath is tuned at $T_i = \frac{ \hbar}{4\pi a_i}$, and the radiation emitted from the bath comes to $r=a_i + \frac{2\sigma_s}{a_i}$ and becomes a constituent of that part of the gravity condensate. Then, the temperature is blueshifted to $\frac{T_i}{\sqrt{1-\frac{a_i}{r_i}}}\sim \frac{m_p}{\sqrt{n_s}} \sim \epsilon_{max}$ (from \eqref{r_a_i}), which means that the left-hand side of \eqref{no_trap} is given by $\frac{\hbar}{\epsilon_{max}} \sim \sqrt{n_s} l_p$. Similarly, the right-hand side can be calculated as $\sqrt{r_i(r_i-a_i)}\sim \sqrt{n_s}l_p$. Therefore, the condition \eqref{no_trap} is saturated.  

Furthermore, this reversible formation is consistent with Bekenstein's idea \ci{Bekenstein}. At each stage, a photon at $T_i=\f{\hbar}{4\pi a_i}$ comes to the gravity condensate of size $R_i\approx a_i+\frac{2\sigma_s}{a_i}$. Since the wavelength $\sim a_i$ and the size $R_i$ are almost the same, the probability for the photon to enter the condensate is roughly one-half, leading to 1 bit of information \ci{Bekenstein}. 
From this and the previous paragaraph, the photon after the approaching \eqref{r_a_i} has the local energy $\e_{max}$ and the 1 bit of information about its existence. Repeating this process, the uniform dense structure like Fig.\ref{f:dense} appears, and the entropy follows the area law \eqref{S_*}. In this sense, the gravity condensate realizes Bekenstein's idea, including the interior description at the level of \eqref{Einstein}. 

Thus, we can conclude from the above arguments that, at least as long as we apply this model by considering a spherical collapsing matter as a collection of many spherical shells with small energy, the condition \eqref{no_trap} naturally holds as a result of the dynamics of \eqref{Einstein}.\footnote{See also the 7th prospect in Sec.\ref{s:conclusion}.} 
Then, regarding the model as a realization of thermodynamically reversible formation, the condition \eqref{no_trap} and hence the bound \eqref{S_B}, is a consequence of the second law of thermodynamics at the level of the semi-classical Einstein equation \eqref{Einstein}. 
However, the model is not complete yet; for example, it only can reproduce the interior metric with $\eta=1$ because it does not incorporate the effects of interaction between constituent quanta \cite{KY1,KY4}. Therefore, further research will be needed to gain a deeper understanding of the origin of the condition \eqref{no_trap}, although it could be expected from the  phenomenological but generic argument below Fig.\ref{f:static} that \eqref{no_trap} should hold at least for highly-excited spherical static configurations.

\subsubsection{Origin of large tangential pressure $\bra T^\theta{}_\theta\ket_*$}\label{s:pressure}
Before closing this section, we discuss the origin of the large tangential pressure $\bra T^\theta{}_\theta\ket_*$ in \eqref{EMT}. Let us imagine that self-gravity brings together a large number of excited modes representing a collapsing matter, and the gravity condensate is formed (as in Fig.\ref{f:pressure}). As the curvature increases, the vacuum fluctuations of the other modes with various angular momenta are induced to produce the quantum pressure as a 4D non-perturbative effect, even if the collapsing matter (such as dust) does not have a classical pressure \cite{KY4} (see Appendix \ref{A:EMT} for a review).\footnote{In the multi-shell model described above, we can use Israel's junction condition and see that a strong surface pressure occurs on each shell due to the 4D energy-momentum conservation and the energy flux of the created particles \cite{KMY}. It can be shown that its continuum version is $\bra T^\theta{}_\theta \ket_* $ \cite{KY4}.} 

This can also be understood through the 4D Weyl anomaly, which is determined by the curvatures $\MR$: $\bra T^\mu{}_\mu \ket \sim \hbar n \MR^2$ \cite{BD}. From \eqref{EMT}, we have $\bra T^\theta{}_\theta\ket_* \approx \f{1}{2} \bra T^\mu{}_\mu \ket_* $ as the leading relation for $r\gg l_p$ and obtain the large pressure from the large curvatures \eqref{RRR} \cite{KY3}.

Because of the pressure, the excited quanta are not concentrated in the center but are distributed throughout the interior, forming the gravity condensate. The curvature remains finite as in \eqref{RRR} self-consistently, the center has only small energy, and no singularity appears. This is the non-perturbative dynamical mechanism to resolve the singularity. 

\section{Maximum Entropy $S_{max}$}\label{s:BH}
Following the derivation in Ref.\ci{Y1}, we show that the maximum entropy $S_{max}\equiv S[g_{\mu\nu}^*;R)$ agrees with the Bekenstein-Hawking formula including the coefficient $1/4$ for any type of matter fields, and then identify the relation of $R$ and $a_0$ from a thermodynamical argument. 

\subsection{Bekenstein-Hawking formula}\label{s:SBH}
\subsubsection{Local temperature}\label{s:T_loc}
We first find the local temperature $T_{loc}^*(r)$. 
We note that the excited quanta composing the part around $r$ of the gravity condensate 
are in radially accelerated motion against self-gravity to stay there as in Fig.\ref{f:dense}. The required acceleration is 
\begin{equation}\label{acc}
\a_*(r)=\f{1}{\sqrt{2\eta^2\s}}+\MO(r^{-1}).
\end{equation}
Here, we have applied the metric \eqref{interior} and the formula of the proper acceleration in the interior metric of \eqref{metric}: $\a(r)\equiv|g_{\mu\nu}\a^\mu\a^\nu|^{\f{1}{2}}=\f{\p_r \log \sqrt{-g_{tt}(r)}}{\sqrt{g_{rr}(r)}}$, $\a^\mu\equiv u^\nu \N_\nu u^\mu$, and $u^\mu \p_\mu=(-g_{tt}(r))^{-\f{1}{2}}\p_t$. Then, we have $\alpha_*(r)^{-1}\sim \sqrt{n}l_p \sim \MR_*(r)^{-1/2}$ (from \eqref{Dr*}), 
and the quanta can be considered accelerating in a locally flat subsystem. Therefore, we can apply the Unruh effect locally \ci{Jacobson,Pad}: 
\begin{equation}\label{T_loc}
T_{loc}^*(r)=\f{\hbar \a_*(r)}{2\pi}=\f{\hbar}{2\pi\sqrt{2\s\eta^2}}+\MO(r^{-1}),
\end{equation}
which is (from \eqref{sigma}) $\sim \e_{max}$ consistent with \eqref{e=T}. Thus, the excited quanta behave typically like a local thermal state in the radial direction at temperature \eqref{T_loc}. 

This is kinematical and can be applied to any type of local degrees of freedom in the metric \eqref{interior}, due to the universality of the Unruh effect. 
In Sec.\ref{s:formation}, on the other hand, we have got $T_{loc}(R_i)\sim \epsilon_{max}$, which derives \eqref{T_loc}  dynamically through particle creation
(see Ref.\ci{Y1} for its explicit derivation). Therefore, the kinematical and dynamical results coincide, and \eqref{T_loc} is robust.

At first glance, the fact that $T_{loc}^*(r)$ is constant might appear to contradict Tolman's law, while its naive application to \eqref{gtt} for $r\ll R$ would lead to $T_{loc}(r)=\f{T_{H}}{\sqrt{-g_{tt}^*(r)}}\sim \f{\hbar}{\sqrt{\s}}e^{\f{R^2}{4\s\eta}}\gg \frac{m_p}{\sqrt{n}}$, violating the semi-classical condition \eqref{e_max}. In general, Tolman's law holds only if, in a stationary spacetime, thermal radiation (more generally, energy flow) can propagate between objects that are stationary with respect to each other within the considered time, including (if any) the effects of interaction with other modes and scattering by potentials \ci{Tolman,Landau_SM}. On the other hand, using the metric \eqref{interior} with \eqref{gtt}, the radial null geodesic equation is given by $\f{dr(t)}{dt}=\pm\sqrt{K}\f{2\s}{r^2} e^{-\f{R^2-r^2}{4\s\eta}}$, which means that in the absence of interaction, it takes time $\D t \sim e^{a_0^2/l_p^2}$ for thermal radiation to travel a distance $\D r =\MO(a_0)$ inside.

Therefore, we can understand \eqref{T_loc} as follows. At each stage of the adiabatic formation in Sec.\ref{s:formation}, the local temperature of the outermost layer is $T_{loc}(r)\sim \e_{max}$. After the subsequent radiation (or “shell”) covers this part, the local temperature is kept through the redshift.
Only during $\D t < \MO(e^{a_0^2/l_p^2})$ can the gravity condensate exist consistent with Tolman's law. \textit{In this sense}, the entropy-maximizing configuration is not in global equilibrium but in radially local one \footnote{This non-global equilibrium makes the difference from the result of Ref.\ci{Oppenheim}, and is consistent with the self-consistent state $|\psi\ket_*$ (mentioned above \eqref{sigma}). See Prospect 3 in Sec.\ref{s:conclusion} for equilibrium states in quantum gravity.}. This is consistent with the result that Tolman's law does not lead to maximum entropy (see Appendix \ref{A:Tolman}). 
Furthermore, we will see in Sec.\ref{A:Bousso2} that the saturation of a local Bousso bound is realized by the Unruh temperature, not by the Tolman's one.  

\subsubsection{Derivation of the entropy-area law}\label{s:D_S}
Now, in this local equilibrium, the 1D Gibbs relation
\begin{equation}
T_{loc} s = \rho_{1d}+p_{1d}
\end{equation}
holds for $\rho_{1d}=4\pi r^2 \bra - T^{\hat t}{}_{\hat t}\ket$ and $p_{1d}=4\pi r^2 \bra T^{\hat r}{}_{\hat r}\ket$, since the configuration is uniform in the proper radial length $\hat r$ \ci{Groot}. Also, from \eqref{EMT}, $p_{1d}^*=\f{2-\eta}{\eta}\rho_{1d}^*$ plays the role of the equation of state because $\bra T^\theta{}_\theta\ket_*$ comes from the vacuum and thus has no thermodynamic contribution. Indeed, this treatment is consistent with the Bousso bound (see Sec.\ref{A:Bousso2}). 
These together with \eqref{T_loc} and \eqref{EMT} provide
\begin{equation}\label{s2}
s_*(r)=\f{\rho_{1d}^*+p_{1d}^*}{T_{loc}^*}=\f{1}{T_{loc}^*}\f{2}{\eta}\rho_{1d}^*=\f{2\pi \sqrt{2\s}}{l_p^2},
\end{equation}
which is $\sim \f{\sqrt{n}}{l_p}$ and consistent with \eqref{s*}. Applying this and \eqref{interior} to \eqref{S}, we obtain the Bekenstein-Hawking formula \ci{Bekenstein,Hawking}:
\begin{equation}\label{S_BH}
S_{max}=\int^R_{\sim \sqrt{n}l_p}dr \sqrt{g^*_{rr}}s_*=
\int^R_{\sim\sqrt{n}l_p} dr \sqrt{\f{r^2}{2\s}} \f{2\pi \sqrt{2\s}}{l_p^2}\approx\f{{\cal A}}{4l_p^2},
\end{equation}
where ${\cal A}\equiv 4\pi R^2 = 4\pi a_0^2+\MO(1)$. This gives the precise version of \eqref{S_*}, and therefore \eqref{S_B} and \eqref{S_BH} mean \eqref{S_bound}. 

Note that while the entropy density \eqref{s2} depends on the species of fields and the equation of state 
through $\s$ in \eqref{sigma}, 
$\s$ cancels out in \eqref{S_BH}, leading to the coefficient $1/4$. 
This is the result of the self-consistent  self-gravity \eqref{interior} and the universal temperature \eqref{T_loc} \ci{Y1}.

\subsection{Surface}\lb{s:surface}
We here determine the relation of the size $R$ and the total energy $M_0=\frac{a_0}{2G}$ from thermodynamics \ci{Y1}. Imagine that the gravity condensate is in equilibrium with a heat bath of temperature $T_0$ during $\D t < \MO(e^{a_0^2/l_p^2})$.  
From thermodynamic relation $T_0dS=dM_0$ and \eqref{S_BH}, the equilibrium temperature is determined as the Hawking temperature \cite{GH}: 
\begin{equation}\label{T_H}
T_0=\f{\hbar}{4\pi a_{0}}.
\end{equation}

Now, radiation emitted from the bath at $r\gg a_0$ comes close to the surface at $r=R$, and according to Tolman's law in the Schwarzschild metric, the blueshifted temperature is given by $\f{T_0}{\sqrt{1-\f{a_{0}}{R}}}$. On the other hand, the argument based on the internal structure determines the local temperature at $r\leq R$ as \eqref{T_loc}. Then, considering the interior and exterior parts as two thermodynamic phases, thermodynamics requires that the local temperature be continuous at the boundary, $r=R$ \ci{Landau_SM}:\footnote{Here, the energy flow is balanced between the condensate and the heat bath, and there is no net energy flow. Therefore, the latent heat at the boundary should be zero, and the local temperature should be continuous \cite{Nakagawa-Sasa}.
} 
\begin{equation}\label{T_cond}
T_{loc}^*(R)=\f{T_0}{\sqrt{1-\f{a_0}{R}}}.
\end{equation}

This determines the relation of $R$ and $a_0$. Setting $R=a_{0}+\D$ ($\D\ll a_{0}$) and using \eqref{T_loc} and \eqref{T_H}, \eqref{T_cond} becomes $\f{\hbar}{2\pi \sqrt{2\s\eta^2}}=\f{\hbar}{4\pi a_{0}}\f{1}{\sqrt{\f{\D}{R}}}\approx \f{\hbar}{4\pi \sqrt{a_{0}\D}}$, leading to $\D=\f{\s\eta^2}{2a_0}$. We thus obtain \eqref{size0}:
\begin{equation}\lb{size}
R=a_{0}+\f{\s\eta^2}{2a_{0}}.
\end{equation}
This satisfies Israel's condition and can also be obtained from a mechanical argument \cite{Y1}. Note that the proper length $\sqrt{g_{rr}^*(R)}\f{\s\eta^2}{2a_{0}} \sim \sqrt{n}l_p~(\gg l_p)$ is independent of $a_0$. 

Thus, using \eqref{size} and fixing $A_0$ with the same procedure to get \eqref{gtt}, we obtain the complete form of $g_{\mu\nu}^*$:
\begin{equation}\label{interior2}
ds^2=-\f{\s\eta^2}{2r^2}e^{-\f{R^2-r^2}{2\s\eta}}dt^2+\f{r^2}{2\s}dr^2+r^2d\Omega^2,
\end{equation}
which is \eqref{interior0}.

We finally summarize the thermodynamic properties of the gravity condensate. It has the Bekenstein-Hawking entropy \eqref{S_BH} and the Hawking temperature \eqref{T_H}, which leads to the negative specific heat: $C_*=T_0 \frac{dS}{d T_0} = -2 \frac{\pi a_0^2}{l_p^2}<0$.\footnote{A small subsystem where self-gravity is negligible has the entropy density $s_*(r)\sim \frac{\sqrt{n}}{l_p}$ (from \eqref{s2}) and the local temperature $T_{loc}^*(r)\sim \frac{\hbar}{\sqrt{n} l_p}$ (from \eqref{T_loc}), providing the relation $s_* \sim n T_{loc}^*$. Therefore, the specific heat is positive in the local subsystem \cite{Landau_SM}. Considering the whole system where self-gravity is involved, however, the total specific heat $C_*$ is negative, as in the case of self-gravitating thermal radiation (see Appendix \ref{A:radiation}).} These are the same as the standard ones \cite{Hawking,BD}, and thus the gravity condensate is consistent. 

\section{Bousso bound}\label{s:Bousso}
In this section, we use the results obtained so far to derive the Bousso bound. We then confirm that the gravity condensate saturates not only the Bousso bound but also the \textit{local} sufficient conditions proposed in the literature.
\subsection{Derivation of the Bousso bound}
\begin{figure}[h]
\begin{center}
\includegraphics*[scale=0.5]{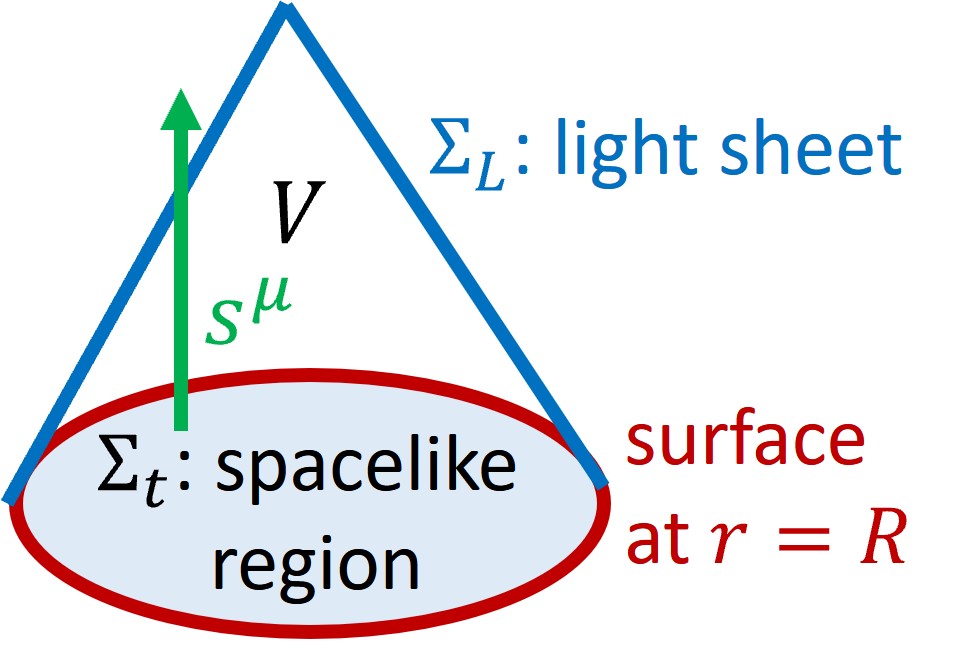}
\caption{A spacetime region $V$ enclosed with $\Sigma_t$ and $\Sigma_L$.}
\label{f:light}
\end{center}
\end{figure}

\lb{s:Ver_B}
We show that the Bousso bound holds in our class of configurations. We first check that the entropy evaluated on a $t$-constant spacelike hypersurface $\Sigma_t$ inside the surface at $r=R$, which we have considered so far, agrees with one evaluated on a spherical ingoing null hypersurface $\Sigma_L$ starting from the surface at $r=R$ and converging at $r=0$ (a light sheet) \cite{Bousso1}. We suppose here that all configurations in our class satisfy \eqref{Einstein} and are regular.\footnote{At lease, we have seen in Sec.\ref{s:No_sing} that the entropy-maximized, most dense, configuration is regular.} 
Applying $\N_\mu \tilde s^\mu=0$ to a finite spacetime region $V$ 
enclosed with $\Sigma_t$ and $\Sigma_L$ ($\p V=\Sigma_L-\Sigma_t$ as in Fig.\ref{f:light}) 
and using Gauss' theorem, 
we obtain
\begin{equation}\lb{Gauss}
0 =\int_V d {\mathcal V}\N_\mu \tilde s^\mu = \int_{\p V} d \Sigma_\mu \tilde s^\mu 
\Rightarrow \int_{\Sigma_t} d \Sigma_\mu \tilde s^\mu = \int_{\Sigma_L} d \Sigma_\mu \tilde s^\mu. 
\end{equation}
Therefore, the entropy $S[g_{\mu\nu},R)$, \eqref{S_typ}, agrees with the covariant entropy. 
Then, combining this and \eqref{S_bound}, we obtain the Bousso bound for thermodynamic entropy. 

Thus, from this result and the discussion so far,
we conclude that in the class of spherically-symmetric static configurations, only the solution metric \eqref{interior2} saturates the Bousso bound. 

\subsection{Local sufficient conditions for the Bousso bound}\lb{s:Suf_B}
We study the relation of our discussion and
the sufficient conditions proposed in Refs.\ci{FMW,BFM} for the Bousso bound 
and check the consistency. 

First, we prepare ingredients to be used below. In a static spherical configuration with metric \eqref{metric}, the relation of entropy density $s$ and entropy current $\tilde s^\mu$ is as follows. For a $t$-constant hypersurface, we have $d \Sigma_\mu=-u_\mu \sqrt{g_{rr}}r^2 \sin \theta dr d\theta d\phi$ with $u_\mu dx^\mu=-\sqrt{-g_{tt}}dt$ \ci{Poisson}. Then, \eqref{S} gives $s=4\pi r^2 (-u_\mu \tilde s^\mu)$, i.e.
\begin{equation}\label{ss}
s(r)=4\pi r^2 \sqrt{-g_{tt}(r)}\tilde s^t(r),
\end{equation}
and the entropy current is given by $\tilde s^\mu \p_\mu=\tilde s^t \p_t$.

We consider a spherical light sheet. 
A radially-ingoing null vector $k^\mu$ tangent to a geodesic generating it 
satisfies $0=g_{\mu\nu}k^\mu k^\nu=g_{tt}(k^t)^2+g_{rr}(k^r)^2$, leading to
\begin{equation}\label{kk}
\l(\f{k^t}{k^r}\r)^2=\f{g_{rr}}{-g_{tt}}.
\end{equation}
Using this, we can calculate for a spherical static energy-momentum tensor $\bra T_{\mu\nu}(r)\ket$
\begin{align}\lb{Tkk}
\bra T_{\mu\nu} \ket k^\mu k^\nu&=\bra T_{tt}\ket (k^t)^2+\bra T_{rr}\ket (k^r)^2  \nn\\
 &= g_{rr}(k^r)^2 \l(\l(\f{k^t}{k^r}\r)^2\f{-g_{tt}}{g_{rr}}\bra-T^t{}_t\ket+\bra T^r{}_r\ket\r)\nonumber \\
 &= g_{rr}|k^r|^2 \l(\bra-T^t{}_t\ket+\bra T^r{}_r\ket\r).
\end{align}

\subsubsection{Meaning of the upper bound \eqref{no_trap_Bek}}\label{A:Bousso1} 
A local sufficient condition, Eq.(1.9) in Ref.\ci{FMW} (or one in Ref.\ci{Pesci}), is given by 
\begin{equation}\label{FMW3}
|-\tilde s^\mu k_\mu|\lesssim \f{1}{\hbar} \Delta \zeta \bra T_{\mu\nu}\ket  k^\mu k^\nu,
\end{equation}
where $\D \zeta$ is the (finite) affine length of a geodesic along $k^\mu$ generating a light sheet 
in which quanta contributing to the entropy are included completely. \eqref{FMW3} is a kind of light ray equivalent of Bekenstein's bound \eqref{S_BB}, since $\Delta \zeta$ and $\bra T_{\mu\nu}\ket  k^\mu k^\nu$ correspond to $R$ and $M_0$, respectively.

We here show that the static condition \eqref{no_trap_Bek} (leading to the bound \eqref{S_B}) is a static spherical version of \eqref{FMW3}.
First, we apply \eqref{ss} to \eqref{no_trap_Bek} and get 
\begin{equation}\lb{FMW1}
\tilde s^t \lesssim \f{1}{\hbar}\sqrt{\f{g_{rr}}{-g_{tt}}}(r-a(r))\bra -T^t{}_t\ket.
\end{equation}
We here have $|-\tilde s^\mu k_\mu|=|-\tilde s^t k_t|=\tilde s^t(-g_{tt})|k^t|$. 
Also, for excitations consistent with thermodynamics, 
$\bra-T^t{}_t\ket\gtrsim\bra T^r{}_r\ket$ holds \ci{Landau_SM}, 
and \eqref{Tkk} means 
$\bra T_{\mu\nu} \ket k^\mu k^\nu \sim g_{rr}|k^r|^2 \bra-T^t{}_t\ket$. 
Using these, \eqref{FMW1} leads to 
\begin{align}\lb{FMW2}
|-\tilde s^\mu k_\mu| &\lesssim \f{1}{\hbar}\sqrt{\f{-g_{tt}}{g_{rr}}} \f{|k^t|}{|k^r|} \f{r-a(r)}{|k^r|} \bra T_{\mu\nu} \ket k^\mu k^\nu \nn\\
 &= \f{1}{\hbar} \f{r-a(r)}{|k^r|} \bra T_{\mu\nu}\ket k^\mu k^\nu,
\end{align}
where \eqref{kk} is applied.
In terms of the affine parameter $\zeta$ of the geodesic, 
we have $k^r=\f{dr(\zeta)}{d\zeta}$ and express $\f{r-a(r)}{|k^r|}=\l|\f{d\zeta}{dr}\r| (r-a(r)) = \Delta \zeta$. 
Thus, \eqref{FMW2} provides \eqref{FMW3} at each $r$ along the geodesic.  

We discuss the meaning of this $\Delta \zeta$. 
In general, when considering a light sheet going to a horizon, 
the sheet must not be continued to the interior, 
since the horizon entropy contains the contribution from objects that fell inside: otherwise, it would be counted twice \ci{Bousso2}. 
In the geometry \eqref{metric} without trapped surface, 
the ``would-be horizon" for a point $r$ exists at $r=a(r)$. 
Therefore, our $\D \zeta$ represents the affine distance from a point to the ``would-be end point", and hence our $\D \zeta$ corresponds to that of Eq.(1.9) in Ref.\ci{FMW}.

Thus, we conclude from the arguments so far that the spacetime $g_{\mu\nu}^*$ \eqref{interior2} is derived as a necessary condition for saturating \eqref{FMW3}. 

\subsubsection{Exact saturation of a local sufficient condition by $g_{\mu\nu}^*$}\label{A:Bousso2}
We study another local sufficient condition, Eq.(3.5) in Ref.\ci{BFM}:  
\begin{equation}\label{BFM}
|k^\mu k^\nu \N_\mu \tilde s_\nu| \leq \f{2\pi}{\hbar} \bra T_{\mu\nu} \ket k^\mu k^\nu,
\end{equation}
which is motivated by an argument on the non-locality of entropy. We show that the gravity condensate with \eqref{interior2} saturates this exactly. 

We first construct the spherically-symmetric static version of \eqref{BFM}. 
From \eqref{ss}, we have $\tilde s_\mu dx^\mu=\tilde s_tdt$ with $\tilde s_t=-\f{\sqrt{-g_{tt}}}{4\pi r^2}s$. 
We can calculate in \eqref{metric}
\begin{align}\lb{kks}
&k^\mu k^\nu \N_\mu \tilde s_\nu \nn\\
&=k^t k^r (\N_t \tilde s_r + \N_r \tilde s_t)\nn\\
&=k^t k^r \sqrt{-g_{tt}} \l(\f{s}{4\pi r^2} \p_r \log \sqrt{-g_{tt}} - \p_r\l(\f{s}{4\pi r^2}\r)\r).
\end{align}
From this and \eqref{Tkk}, \eqref{BFM} becomes 
\begin{align}
&\bra-T^t{}_t\ket+\bra T^r{}_r\ket \nn \\
&\geq \f{\hbar}{2\pi}\f{\sqrt{-g_{tt}}}{g_{rr}} \f{|k^t|}{|k^r|} 
\l|\f{s}{4\pi r^2} \p_r \log \sqrt{-g_{tt}} - \p_r\l(\f{s}{4\pi r^2}\r)\r|\nn\\ 
&=\f{\hbar}{2\pi}  \l|\f{\p_r \log \sqrt{-g_{tt}}} {\sqrt{g_{rr}} } \f{s}{4\pi r^2} 
-\f{1}{\sqrt{g_{rr}}} \p_r\l(\f{s}{4\pi r^2}\r)\r|
\end{align}
where we have applied \eqref{kk}.
We then use the acceleration $\a=\f{\p_r \log \sqrt{-g_{tt}}} {\sqrt{g_{rr}} }$ to obtain the spherically-symmetric static version of \eqref{BFM}: 
\begin{equation}\label{BFM_cond}
\l|\f{\hbar \a}{2\pi} \f{s}{4\pi r^2} - \f{\hbar}{2\pi} \f{1}{\sqrt{g_{rr}}}\p_r \l(\f{s}{4\pi r^2}\r)\r|
\leq \bra -T^t{}_t\ket + \bra T^r{}_r\ket.
\end{equation}
Note that the entropy bound in this special case involves only the combination $\bra -T^t{}_t\ket + \bra T^r{}_r\ket$, independently of $\bra T^\theta{}_\theta\ket$, and our evaluation of the entropy density in Sec.\ref{s:D_S} is consistent. 

The interior metric \eqref{interior2} saturates \eqref{BFM_cond} exactly at the leading order for $r\gg l_p$: The left-hand side becomes $\f{\hbar \a_*}{2\pi} \f{s_*}{4\pi r^2} - \f{\hbar}{2\pi} \f{1}{\sqrt{g^*_{rr}}}\p_r \l(\f{s_*}{4\pi r^2}\r) \approx \f{\hbar \a_*}{2\pi} \f{s_*}{4\pi r^2} = \f{1}{4\pi G\eta r^2}$ for $r\gg l_p$ (from the local temperature \eqref{T_loc} and the entropy density \eqref{s2}), while the right one becomes $\bra -T^t{}_t\ket_* + \bra T^r{}_r\ket_*=\f{1}{4\pi G\eta r^2}$ (from \eqref{EMT}). Thus, the gravity condensate 
saturates not only the global bound \eqref{S_bound} but also the local condition \eqref{BFM} relevant to the local structure. This is so non-trivial that there must be something behind it.

\section{10 future prospects}\label{s:conclusion}
The identity of black holes is still unknown, and the microscopic constituents should involve fundamental degrees of freedom in quantum gravity, where there is no notion of classical spacetime geometry. Motivated by thermodynamics and (local) holography, in the present paper, we adopted the characterization that a black hole maximizes thermodynamic entropy for a given surface area. For highly-excited spherical static configurations where the condition \eqref{no_trap} holds, we uniquely obtained the entropy-maximizing configuration consistent with local thermodynamics and the 4D semi-classical Einstein equation with many matter fields. That is, a self-gravitating collection of near-Planckian excited quanta condensates into the radially-uniform dense configuration without horizons or singularities as in Fig.\ref{f:dense}, where the self-gravity and large quantum pressure are balanced (called \textit{semi-classical gravity condensate}). The interior metric \eqref{interior2} (say, for $n_s$ scalar fields, with the parameters \eqref{sigma}) satisfies the semi-classical Einstein equation self-consistently and non-perturbatively in $\hbar$. The bulk dynamics makes the entropy of the interior quanta exactly follow the Bekenstein-Hawking formula, leading to the Bousso bound for thermodynamic entropy and the saturation of the local sufficient conditions. This is a candidate picture of black holes in quantum theory. 



Let us discuss the ten future prospects and speculate on the possibility of a more complete quantum description of black holes.

\vskip\baselineskip
\textbf{\textit{1. Role of self-gravity in holography---.}} As shown in Sec.\ref{s:D_S}, the self-gravity, represented by the self-consistent metric \eqref{interior}, changes the entropy \eqref{S} from the volume law to the area law \cite{Y1}. Then, what happens to quantum fields in the metric? From \eqref{s2}, the entropy per unit proper volume is given by $\f{s_*(r)}{4\pi r^2}=\MO\l(\f{\sqrt{n}}{l_p r^2}\r)$. This is much smaller than the naive estimate without the effect of self-gravity, $\MO\l(\f{n}{l_p^3}\r)$, where all modes are assumed to be excited \cite{Bousso2}. This gap should be related to the self-consistent state $|\psi\ket_*$ (described above \eqref{sigma} and \eqref{psi_EMT}) in which only s-waves are excited and the other modes not. Indeed, the number of possible patterns of excited s-waves reproduces the Bekenstein-Hawking formula at a WKB-approximation level \cite{KY4}. Therefore, we can conjecture that \textit{the self-gravity suppresses the excitation of the local degrees freedom in the bulk and reduces the number of active bulk degrees of freedom, leading to the holographic property of entropy.} This should be related to the remarkable fact that the local sufficient condition \eqref{BFM} for the Bousso bound is exactly saturated by the metric \eqref{interior}, which could be an essence of the bulk dynamics consistent with the holographic principle. For that, one could study quantum fields in the metric \eqref{interior} by a semi-classical/perturbative Wheeler-De Witt equation \cite{Kiefer,Kuchar,Suvrat}, since the Hamiltonian constraint $\MH=0$ plays the key role in obtaining $S[g_{\mu\nu};R)$ in Sec.\ref{s:S[g]}. 

\vskip\baselineskip
\textbf{\textit{2. Relation to other gravity-condensate models---.}} We discuss the relation to other gravity-condensate models and explore a possible path to a full-quantum formulation of black holes. 

A model is obtained in the framework of group field theory, a second quantization of loop quantum gravity \cite{Oriti1}. Gluing spherically-symmetric quantum-gravity states kinematically and maximizing the entropy for a given surface area, a quantum gravitational configuration appears with the entropy proportional to the surface area, and the coefficient is fixed by using the Unruh effect and thermodynamic relations. A remarkable point is that the holographic property of the entropy of the quanta living in the interior bulk holds for any size $r$, which corresponds to the radial uniformity in our semi-classical gravity condensate. (See Sec.8 of Ref.\cite{Y1} for more discussions.)

Another one is a view of black holes as Bose-Einstein condensates of gravitons \cite{Dvali3}. Introducing the occupation number $\MN$ of gravitons in a gravitational field of a source with mass $M_0=\frac{a_0}{2G}$ and size $R$ by $\MN=\frac{M_0 a_0}{\hbar} \sim \frac{a_0^2}{l_p^2} (\leq \frac{R^2}{l_p^2})$ and maximizing it for a given size $R$, 
the black hole is characterized by $\MN_{max}\sim \frac{R^2}{l_p^2}$. 

To examine the relation to our gravity condensate, we first review how to obtain $\MN \sim \frac{M_0 a_0}{\hbar}$ \cite{Dvali3}. In the approximation of linear gravity, the gravitational energy of a source with size $R$ and mass $M_0$ can be estimated by $E_g\sim \frac{G M^2_0}{R}$, while the characteristic energy of a single graviton is given by $\e_g\sim\frac{\hbar}{R}$. Therefore, we obtain the occupation number of gravitons as $\MN\sim \frac{E_g}{\e_g}\sim \frac{M_0 a_0}{\hbar}$. 

In the picture of Fig.\ref{f:subsystem}, we consider a spherical subsystem with a width $\Delta \hat r$ and the local energy $\Delta E_{loc}(r)=4\pi r^2 \bra T^{\hat t \hat t}(r)\ket \Delta \hat r$. For the condition \eqref{WKB}, the approximation of linear gravity is valid within the subsystem. Following the above idea, then the gravitational energy can be estimated as $\Delta E_g (r) \sim \frac{G \Delta E_{loc}(r)^2}{\Delta \hat r}$, and the characteristic energy of a single graviton in the subsystem is given by $\epsilon_g (r)\sim \frac{\hbar}{\Delta \hat r}$. 
Therefore, the occupation number of gravitons within the subsystem is estimated by 
\begin{equation}\label{Ng}
N_g(r)\sim \frac{\Delta E_g(r)}{\e_g(r)}\sim \frac{G \Delta E_{loc}(r)^2}{\hbar }. 
\end{equation}
In particular, for the case of our gravity condensate, we have $\Delta E_{loc}^*(r)\sim \sqrt{n}m_p$ (see around \eqref{a*_range}) and get 
\begin{equation}\label{Ng*}
N_g^*(r)\sim n.     
\end{equation}

Here, interesting similarities can be observed between $n$ and $\MN$. For our condensate, the temperature $T_{loc}^*(r)\sim \frac{m_p}{\sqrt{n}}$, characteristic wavelength $\lambda_*(r)\sim \sqrt{n}l_p$, and energy $\Delta E_{loc}^*(r)\sim \sqrt{n}m_p$ are controlled by $n$. These manifestations of $n$ are exactly the same as those of $\MN$ in the condensate of Ref.\cite{Dvali3}, although the former is for local subsystems and the latter for the entire system. 
Also, from \eqref{Ng*}, the condition $n\gg1$ corresponds to that for the classicality in Ref.\cite{Dvali3}. Therefore, a possibility is that realizing the condensate in Ref.\cite{Dvali3} in local subsystems and connecting them consistently with the semi-classical Einstein equation lead to our gravity condensate. 

On the other hand, we have introduced the occupation number of excited matter quanta in the subsystem by \eqref{N}: $N(r)\sim \frac{\Delta E_{loc}}{\epsilon(r)}$. For generic configurations, $N(r) \neq N_g(r)$ holds, while our gravity condensate satisfies 
\begin{equation}\label{NnN}
    N^*(r)\sim n \sim N_g^*(r)
\end{equation}
from \eqref{N*} and \eqref{Ng*}.  This implies that matter quantum and gravity quantum in our gravity condensate play the same role in a sense, or they are indistinguishable as a quantum of some degree of freedom in quantum gravity. Then, \textit{the semi-classical gravity condensate would be a mixture of matter and gravity quanta, providing a basis of exploration of new degrees of freedom.}

Note also that the two other models above discuss quantum gravitational effects without considering quantum matter ones, while our model describes matter quanta in the self-consistent (and non-perturbative) classical gravitational field. 
Therefore, a more detailed study of the connection between the three models would provide some clues to the above expectation. 

\vskip\baselineskip
\textbf{\textit{3. Gravity-condensate phase---.}} 
We note that the gravity condensate, despite being self-gravitational, is uniform in the radial direction: (the leading values of) the local temperature $T_{loc}(r)$, the entropy density $s(r)$, and the 1D energy density $\rho_{1d}(r)$ are constant (see Sec.\ref{s:D_S}). This implies \cite{Landau_SM} that the gravity condensate is a kind of thermodynamic phase. (Let us call it the \textit{gravity-condensate phase}). Indeed, this view works well in determining the position of the surface \eqref{size}. 

The phase could be \textit{quantum gravitational}. 

In general, for materials without self-gravity, quantum effects dominate the determination of macroscopic properties 
when the thermal wavelength $\lambda_T=\frac{\hbar}{\sqrt{m T}}$, a quantum length scale, is of the same order as 
the mean inter-particle distance $\rho_N^{-1/3}$, a classical length scale \cite{Landau_SM2}. Here, $m,~T,~\rho_N$ are mass of particles, temperature, and number density, respectively. 

On the other hand, the relation (from \eqref{WKB}) 
\begin{equation}\label{CG_con}
\lambda(r)\sim \f{\hbar}{\e(r)}\lesssim 
\MR(r)^{-\f{1}{2}} 
\end{equation}
could be regarded as a condition under which the semi-classical approximation \eqref{Einstein} holds: only when quanta have wavelengths $\lambda(r)$ shorter than the radius of spacetime curvatures $\MR(r)^{-\f{1}{2}}$ in the self-consistent gravitational field should the locality of quanta and spacetime be well-defined (at the resolution of $\sim \lambda(r)$) and the concept of the classical and continuum spacetime be established (as in Fig.\ref{f:subsystem}). 

Here, noting the typical relation \eqref{e=T}, we have a relativistic thermal wavelength $\lambda(r)\sim \frac{\hbar}{T_{loc}(r)}$. Therefore, we could expect the correspondence:  
\begin{equation}\label{QG_con}
    \lambda_T \sim \rho_N^{-\frac{1}{3}}
       \overset{?}{\longleftrightarrow}
        \lambda(r) \sim \MR(r)^{-\f{1}{2}}.
\end{equation}

For the gravity condensate, \eqref{CG_con} is saturated: $\lambda_*(r) \sim \MR_*(r)^{-1/2}$ (from \eqref{Dr*}), 
and also the equivalence in the number of gravity and matter quanta \eqref{NnN} holds. Thus, the correspondence \eqref{QG_con} would imply that \textit{the gravity condensate phase is governed essentially by quantum-gravitational effects, but due to the large occupation number \eqref{NnN} for $n\gg1$, the mean-field approximation of quantum gravity by \eqref{Einstein} \cite{Kiefer} holds albeit barely, leading to the semi-classical description \eqref{interior2}.} This could correspond to a quantum liquid of gravity and matter.

A way to study this speculation is to first find an appropriate order parameter, study a Ginzburg-Landau-like theory or Gross-Pitaevskii-like equation including the effect of the redshift, and reproduce the radial uniformity or the interior metric. One could then consider the quantum many-body model behind it.

We now discuss a notion of global thermodynamic equilibrium in quantum gravity (related to footnote \ref{foot:thermal}). 
In the absence of self-gravity, the most typical configurations maximize the entropy for given macroscopic parameters and correspond to the global thermodynamic equilibrium states. If this is also the case for quantum gravity, and if gravity condensates are quantum gravitational objects, then from the derivation procedure, \textit{the global thermodynamic equilibrium states in quantum gravity for a fixed surface area would be gravity condensates}.

The key property in this equilibrium state is that the redshift $g_{tt}(r)\sim - e^{-\frac{R^2-r^2}{2\eta \sigma}}$ is so strong that Tolman's law does not hold (as seen in Sec.\ref{s:T_loc} and Appendix \ref{A:Tolman}). To understand the origin more deeply, the effect of interactions should be important. In general, a phase is a uniform equilibrium state which is achieved through internal interactions \ci{Landau_SM}.
Indeed, we can use a simple model and confirm that such interactions equilibrates the system in a heat bath \ci{KY2}. We could thus expect that due to many such interactions inside, energy should be transferred as \textit{conductive heat rather than radiative heat}, and the propagation timescale should be consistent with $\D t \sim \MO(e^{a_0^2/l_p^2})$ (see Sec.\ref{s:T_loc}) so that the gravity condensate exists as a phase.

\vskip\baselineskip
\textbf{\textit{4. Path-integral evaluation of $S[g_{\mu\nu};R)$---.}} We have assumed the phenomenological form of thermodynamic entropy \eqref{S} for highly excited states and utilized local typicality and the Hamiltonian constraint to estimate the entropy $S[g_{\mu\nu};R)$, \eqref{S_typ}. This is a rough estimate but is the first one (in our knowledge) that includes non-perturbatively the effect of self-gravity consistent with the semi-classical Einstein equation for various configurations. However, this should be justified in a more microscopic manner. 

A way to evaluate $S[g_{\mu\nu};R)$ field-theoretically is a path-integral method \cite{BY1, BY2}. For massless scalar fields in a self-consistent configuration $(g_{\mu\nu},|\psi\ket)$, we should be able to use the propagator of massless particles restricted to a given size $R$ in the metric $g_{\mu\nu}$ and evaluate the density of states. 

In the context of quantum gravity, a similar problem has been studied \cite{Jacobson_vol}; the dimension of the Hilbert space of a spatial region with a fixed proper volume is evaluated in the leading order saddle point approximation, where only gravity contribution is considered, to obtain the entropy-area law associated with the surface area of the saddle ball (see \cite{Bianca} for a discrete model). On the other hand, motivated by (local) holography, we have fixed a surface area and considered typical configurations satisfying the semi-classical Einstein equation with matter fields, to find the entropy-maximizing one with the metric \eqref{interior2}, leading to the Bekenstein-Hawking formula. Therefore, to understand the relation of the two clearly, one would consider a path integral in both gravity and matter for a fixed surface area and examine the relation between the saddle points and the self-consistent solutions of the semi-classical Einstein equation. 

\vskip\baselineskip
\textbf{\textit{5. Thermodynamic entropy vs entanglement entropy---.}} Another field-theoretic approach to $S[g_{\mu\nu};R)$ would be to formulate it as entanglement entropy. 
In such cases, the Unruh temperature is often applied to local Rindler regions as the local temperature \ci{Minic,Casini,Jacobson_entangle}, but it does not necessarily agree with the thermodynamic temperature consistent with the self-consistent gravity. In the case of, say, self-gravitating thermal radiation, 
the Unruh temperature is different from the local temperature obtained by applying Stefan-Boltzmann law or Tolman's law in the metric (see Appendix \ref{A:radiation}). This is natural because the local temperature of an object in global thermodynamic equilibrium, which is fixed by Tolman's law in the self-consistent metric, does \textit{not} generically coincide with the Unruh temperature determined by the acceleration required to stay against the self-gravity. Therefore, the entanglement entropy calculated by such methods can be different from the thermodynamic entropy $S[g_{\mu\nu};R)$. The crucial point is whether to take into account the self-gravity determined by the semi-classical Einstein equation.

As argued in Sec.\ref{s:T_loc}, however, the local temperature \eqref{T_loc} of the gravity condensate can be obtained by applying the Unruh formula locally to the self-consistent metric \eqref{interior2}. 
Therefore, we expect that applying such field-theoretic techniques locally in the metric enables us to evaluate the entropy density of the gravity condensate as entanglement entropy, and the entanglement entropy and the thermodynamic (Boltzmann) one agree and give the Bekenstein-Hawking formula. Indeed, this is the case for the quantum gravity condensate proposed in Ref.\cite{Oriti1}. We would like to check this expectation in the future. 

Note here that it is not clear whether entanglement entropy explains the Bekenstein-Hawking formula in general. As noted in footnote \ref{foot:BH_formula}, the latter is a nontrivial function of gravitational charges that satisfies the first law of thermodynamics, and should be a type of thermodynamic entropy. Entanglement entropy, on the other hand, depends on the quantum state, and its leading value is fixed only by the surface area of the boundary and is not relevant to gravitational charges directly.

\vskip\baselineskip
\textbf{\textit{6. Physical resolution of information problem---.}} 
The purification of the initial Minkowski vacuum after evaporation and the ``formal" derivation of the Page curve are only part of the information problem. More essentially, we need to understand the dynamical mechanism by which the initial wavefunction of a collapsing matter that forms a black hole is recovered after evaporation, leading to the physical Page curve as a result of the unitary evolution of matter and gravity. We also need to clarify the interior structure with quantum dynamics that resolves the singularity. Such overall consistency will reveal the true identity of black holes.

In the gravity condensate, the information of a (typical) collapsing matter is stored in the bulk interior, and the amount agrees with the Bekenstein-Hawking formula. The condensate has neither horizon nor singularity (although we still need to understand the small center part at a fully quantum-gravity level), and almost all parts evaporate in the vacuum due to the Hawking-like radiation. Therefore, it should be natural to expect that (most of) the information recovers after the evaporation. As mentioned above, however, we still need to clarify the mechanism consistent with the energy flow: how the information of the matter leaks out gradually during the evaporation. A possibility is that scattering between a collapsing matter and the radiation occurs frequently at each point inside \cite{KY2}, and such interactions should transfer the information to the emitted radiation, reproducing the physical Page curve.

To see this explicitly, one would at least need to check the initial-state dependence of the radiation emitted after the interaction. However, the condensate is in a typical state, and such dependence is not easy to see. One idea to overcome this is to consider perturbations from the typical state, adding a small atypical portion to the condensate and tracking it during evaporation. Another is to study a protocol that distinguishes between two typical states and implement it in this model.

\vskip\baselineskip
\textbf{\textit{7. Non-typical configurations---.}} 
The gravity condensate, which is the entropy-maximizing configuration for a given surface area, can be obtained by reversible processes in a heat bath, as shown in Sec.\ref{s:formation}. Then, what is non-typical configurations that are formed by a generic collapse of matter? 

One way to discuss this is to study the time evolution of a collapsing matter including the backreaction from particle creation during the collapse, considering the point in footnote \ref{foot:backreaction}. Indeed, we can consider the matter as consisting of many spherical shells with small energy, distributed according to the initial state, and solve \eqref{Einstein} self-consistently \cite{KY2}. It shows that the dense structure with the metric \eqref{interior2} is formed only around the surface (in which \eqref{r_a_i} occurs), while the structure in deeper regions depends on the details of the initial distribution. The configuration has a smaller entropy than \eqref{S_BH}, is atypical and thus is not a  black hole according to our definition.

Here, it is important to consider a finite width even when discussing collapse of a spherical shell.  For simplicity it is often modeled by an infinitely thin shell \cite{BD,BHmodel}, but physically, any collapsing matter is an excitation of quantum fields with a physical information $|\psi \ket$, which due to the non-locality, cannot be localized completely in the radial direction. Considering a finite-width shell as a collection of many tiny shells and including the evaporation effect, we can see that the tangential pressure $\bra T^\theta{}_\theta\ket$ occurs dynamically and the gravity-condensate structure appears around $r=a_0$, while the interior is vacuum \cite{KY2,KY4}. This is a result of the 4D dynamics: the 4D conservation law $\nabla_\mu \bra T^\mu{}_\nu\ket=0$ contains $\bra T^\theta{}_\theta\ket$, which can be large due to the 4D Weyl anomaly \cite{KY3} (as in Sec.\ref{s:pressure}). In the 2D cases, there is no tangential direction.

Another way to answer the question is to consider a solution space, consistent with the 4D Weyl anomaly, of the semi-classical Einstein equation and study various configurations in a non-perturbative manner for $\hbar$ \cite{HKLY}. It shows that the most typical ones for a fixed surface area are similar to the gravity condensate, and the number of such configurations agrees with the Bekenstein-Hawking entropy (except for $O(1)$ numerical factor). Non-typical ones have the dense structure only around the surface, which is consistent with the one above. 

Thus, these different studies based on the 4D non-perturbative dynamics of the semi-classical Einstein equation give almost the same picture of the typical and non-typical configurations. To clarify their relation, in the future, we would like to consider a semi-classical time-dependent perturbation and study how such non-typical ones evolve to the typical one (relaxation process) and how stable the gravity condensate is. 

As an alternative approach to the stability problem, one can consider an information-theoretic version of singularity theorem: Under the assumption of the Bousso bound, singularities form when a spatial region has an entropy greater than the bound \cite{B_singularity}. Therefore, if the Bousso bound is assumed and if it is possible to add some amount of entropy to the gravity condensate in a way consistent with \eqref{Einstein}, it would transform into a usual black hole with singularities, indicating an instability.

\vskip\baselineskip
\textbf{\textit{8. Relation to the classical picture of black holes---.}} One might wonder why the gravity condensate is so different from the classical picture of black holes, a vacuum region surrounded by a horizon. 
The main reason for this difference is that many previous studies have not properly considered 4D quantum dynamics. 
First, note again that the classical one was originally derived from the classical dynamics of a collapsing matter, and that the existence of horizons has not been confirmed observationally yet. At the semi-classical level, as discussed above, infinitely thin-shell models and 2D approximation are often used to lead to almost the same picture as the classical one \cite{BHmodel}. In general, however, 2D and 4D dynamics are different even in 4D spherically symmetric systems: UV divergent structures and energy conservation laws. The former provides the difference in 2D and 4D Weyl anomalies, and the latter the (non-) existence of the tangential pressure. As mentioned above, furthermore, the consideration of a finite width and the backreaction from evaporation during the collapse (consistently with footnote \ref{foot:backreaction}) must be considered. These points make a big difference in a self-consistent analysis of the time evolution of a collapsing matter to find a picture of black holes. Indeed, our argument \cite{KMY,KY1,KY2,KY3,KY4,KY5,Y1} takes these points into account and obtains the picture of the gravity condensate. 

Another aspect in the difference between the conventional picture and the gravity condensate is that they belong to different branches in a solution space of the semi-classical Einstein equation. 
In Ref.\cite{HKLY}, we constructed a self-consistent equation of the energy distribution $\frac{a(r)}{2G}$ including the effect of the 4D Weyl anomaly, and examined the structure of the solution space in a non-perturbative manner for $\hbar$. We then found that there exist two branches: one is perturbative and contains Schwarzschild-like metrics, while the other is non-perturbative and includes the dense solution \eqref{interior}. The point is that, the higher derivative term $\Box R_{\rm RicciScalar}$, which appears generically in the anomaly \cite{BD}, causes transitions between the two branches as $r$ changes. 
Therefore, it would be interesting to study 4D semi-classical time evolution of a collapsing matter including such higher derivative effects and see how general the formation of the gravity condensate is and which one is formed, the conventional one or the gravity condensate.

Another interesting question is: Which is more typical, the conventional one or the gravity condensate, in the sense of greater entropy? In Sec.\ref{s:SBH}, we have shown that the gravity condensate has the Bekenstein-Hawking entropy, including the coefficient $1/4$. Therefore, both have the same entropy and should be equally typical for a fixed surface area. To judge this in more detail, one could, say, calculate a correction term to the area entropy and find out which one is larger \cite{Y3}.

\vskip\baselineskip
\textbf{\textit{9. Gravitational field with finite entropy---.}} We briefly comment on entropy in gravity. An interesting aspect in estimating the entropy $S[g_{\mu\nu};R)$ in Sec.\ref{s:entropy} is that the local typicality and Hamiltonian constraint assign a finite entropy to a gravitational field $g_{\mu\nu}$. This is reminiscent of two things. First, it is similar to the idea that the Einstein equation corresponds to the equation of state in spacetime thermodynamics \cite{Jacobson,Pad}. Second, it should be relevant to the view of entropy as a gravitational charge \cite{Wald,SY} (see Sec.8 of Ref.\cite{Y1} for details). 
These could be relevant each other.

\vskip\baselineskip
\textbf{\textit{10. Phenomenology---.}}
We finally discuss the phenomenological aspect of the gravity condensate. In Ref.\cite{CY}, an investigation of imaging of the gravity condensate showed that, despite the absence of an event horizon, the image is significantly darkened by the strong redshift of \eqref{interior0} and almost identical to the classical black-hole image, giving the consistency with the current data. Furthermore, the intensity around the inner shadow is slightly enhanced when the emission is a bit inside the surface, which may be a future observable prediction for characterizing the condensate. It is also interesting to investigate gravitational waves in this model; in particular, some echo signal can be expected due to the existence of the surface structure \cite{echo}. For a more realistic phenomenology, furthermore, it should be important to generalize the gravity condensate to a rotating case (see Ref.\cite{KY2} for a slowly-rotating case).

\section*{Acknowledgments}
Y.Y. thanks C.Barcelo, F.Becattini, R.Casadio, C.Y.Chen, C.Goeller, T.Harada, C.Kelly, E.Livine, N.Nakagawa, A.Pesci, and Y.Sakatani for inspiring discussions and valuable comments.
Y.Y. is partially supported by Japan Society for the Promotion of Science 
(No.21K13929) and by RIKEN iTHEMS Program. 
\appendix
\section{Self-gravitating thermal radiation}\label{A:radiation}
To demonstrate the self-gravity dependence of entropy explicitly in an example, we provide a review for the entropy of self-gravitating thermal radiation in a different manner from Ref.\cite{Sorkin}. We also check its consistency to $S[g_{\mu\nu};R)$ and the upper bound.

\subsection{Metric}
We consider a spherically-symmetric static configuration of 
self-gravitating ultra-relativistic fluid with size $R$, mass $M_0=\f{a_0}{2G}$, 
and equation of state $\rho=3p$,
where $\rho(r)\equiv \bra -T^t{}_t(r)\ket$ and $p(r)\equiv \bra T^r{}_r(r)\ket=\bra T^\theta{}_\theta(r)\ket$. 
For simplicity, we here neglect a small contribution from Weyl anomaly from the curvatures \ci{BD} and interactions \ci{fluidbook}.
We construct its interior metric, for \eqref{metric}, in a heuristic manner. (See \ci{Weinberg} for another derivation.) 

First, the equation of state $\rho=3p$, equivalent to $\bra T^\mu{}_\mu\ket=0$, 
means that there is no special length scale in the system. 
Therefore, the order of the magnitude of the curvature at $r$ should be $\MR (r) \sim \f{1}{r^2}$. 
Also, $\rho=3p$ indicates $\rho\sim p$, and both $\rho$ and $p$ contribute to the curvature almost equally. 
Thus, from the Einstein equation \eqref{Einstein}, 
we have $\rho(r)\sim \f{1}{G r^2}$, which can be expressed as 
$\rho(r)=\f{1-C}{8\pi G r^2}$
with a $\MO(1)$ constant $C$. 
Applying \eqref{Hami} to this, 
we have
\begin{equation}\lb{a_r_A}
\f{a(r)}{2G}=\f{1-C}{2G}r.
\end{equation}
This must be positive due to radiation excitation and must be smaller than $\f{4r}{9G}$ from Buchdahl's limit \ci{Buchdahl,Weinberg}, which requires 
$\f{1}{9}<C<1$.
Then, the interior metric \eqref{metric} becomes 
\begin{equation}\lb{metricA1}
ds^2=-C e^{A(r)}dt^2+\f{1}{C}dr^2+r^2d\Omega^2.
\end{equation}

Next, $A(r)$ is determined from the ultra-relativistic fluid condition: 
$\bra -T^t{}_t\ket=3 \bra T^r{}_r\ket$ and $\bra T^r{}_r\ket=\bra T^\theta{}_\theta\ket$.
\eqref{metricA1} gives 
\begin{align}
\lb{GttA}
-G^t{}_t &=\frac{1-C}{r^2},\\
\lb{GrrA}
G^r{}_r &=-\frac{1-C}{r^2}+\frac{C A'(r)}{r},\\
\lb{G33A}
G^\theta{}_\theta &= \frac{C}{2 r}A'(r)+\frac{1}{4} C A'(r)^2 +\f{1}{2}C A''(r).
\end{align}
Using these and \eqref{Einstein}, 
$\bra -T^t{}_t(r)\ket=3 \bra T^r{}_r(r)\ket$ leads to 
$A(r)=A_0 + \f{4(1-C)}{3C}\log r$ ($A_0$: constant). 
Applying this $A(r)$, \eqref{GrrA} and \eqref{G33A} to $\bra T^r{}_r(r)\ket=\bra T^\theta{}_\theta(r)\ket$ through \eqref{Einstein},
we have $(1-C)(-4+7C)=0$, 
giving $C=\f{4}{7}$, which is consistent with Buchdahl's limit. 
Then, we get $A(r)=A_0 + \log r$, and \eqref{metricA1} becomes  
\begin{equation}\lb{metricA2}
ds^2=-\f{4}{7}e^{A_0}rdt^2+\f{7}{4}dr^2+r^2d\Omega^2,
\end{equation}
and the energy density is given by \cite{Weinberg}
\begin{equation}\lb{rho_rad}
\rho(r)=\f{3}{56\pi G r^2}.
\end{equation}
Here, the size $R$ is determined by applying \eqref{a_r_A} with $C=\f{4}{7}$ and $a(R)=a_0$: 
\begin{equation}\lb{R_A}
R=\f{7}{3}a_0.
\end{equation}

Finally, we connect \eqref{metricA2} to the Schwarzschild metric with mass $\f{a_0}{2G}$. 
This requires the continuity of $-g_{tt}(r)$ at $r=R$ \ci{Poisson}: 
$-g_{tt}(R)=\f{4}{7}e^{A_0}R=1-\f{a_0}{R}$, leading through \eqref{R_A} to $e^{A_0}=\f{1}{R}$.
Thus, we reach the final form:
\begin{equation}\lb{metricA3}
ds^2=-\f{4}{7}\f{r}{R}dt^2+\f{7}{4}dr^2+r^2d\Omega^2~~~{\rm for}~~l_p\ll r\leq R.
\end{equation}

Note that the curvatures are small for $r\gg l_p$: $R_{\mu\nu\a\b}R^{\mu\nu\a\b}\sim \f{1}{r^4}$. 
Therefore, the contribution from 4D Weyl anomaly \ci{BD} is indeed small for $r\gg l_p$, 
where our approximation is valid.

\subsection{Entropy}
Now, let us evaluate the entropy. 
First, the equation of state $\rho=3p$ and (local) thermodynamics lead to the Stefan-Boltzmann law: 
\begin{equation}\lb{SB1}
\rho(r)=\f{kn_f}{4\hbar^3}T_{loc}(r)^4,
\end{equation}
where $k$ is a $\MO(1)$ numerical constant, and $n_f$ 
is the number of the degrees of freedom of radiation \ci{Landau_SM}. 
This and \eqref{rho_rad} determine the local temperature at $r$: 
\begin{equation}\lb{T_loc_rad}
T_{loc}(r)\sim \f{n_f^{-\f{1}{4}}\hbar}{\sqrt{l_p r}}.
\end{equation}
Note that this agrees with one obtained by applying Tolman's law \ci{Landau_SM} to the metric \eqref{metricA3},
$\f{T_0}{\sqrt{-g_{tt}(r)}}\sim r^{-\f{1}{2}}$, 
and thus the metric \eqref{metricA3} is consistent with thermodynamics and general relativity. From \eqref{e=T}, this gives the characteristic excitation $\epsilon(r) \sim T_{loc}(r)$ for a typical state $|\psi \ket$. 

Using $\rho=3p$, the Gibbs relation $\rho+p=T_{loc} s_{3d}$ \ci{Groot}, \eqref{SB1} and \eqref{T_loc_rad}, 
we obtain the entropy per unit proper volume, $s_{3d}(r)\equiv \f{s(r)}{4\pi r^2}$, as 
\begin{equation}\lb{SB2}
s_{3d}(r)=\f{k n_f}{3\hbar^3}T_{loc}(r)^3\sim \f{n_f^{\f{1}{4}}}{(l_p r)^{\f{3}{2}}}.
\end{equation}

Thus, applying \eqref{SB2} and \eqref{metricA3} to \eqref{S}, we get \ci{Sorkin}
\begin{equation}\lb{S_total_rad}
S=4\pi\int^R_0dr \sqrt{g_{rr}(r)}r^2 s_{3d}(r)
\sim \int^R_0 dr \f{n_f^{\f{1}{4}}}{l_p^{\f{3}{2}}}r^{\f{1}{2}}\sim n_f^{\f{1}{4}} \l(\f{R}{l_p}\r)^{\f{3}{2}},
\end{equation}
which is different from the usual volume law $\sim R^3$. 
This is a consequence of the self-gravity dependence of entropy.

\subsection{Consistency check}
We use this example to examine the consistency of our argument on $S[g_{\mu\nu};R)$ and the upper bound. 

First, the ratio between \eqref{SB1} and \eqref{SB2} leads to  
\begin{equation}\label{SB_radiation}
    s_{3d}(r) =\f{4\bra -T^t{}_t(r)\ket}{3T_{loc}(r)},
\end{equation}
which gives a consistency to the estimation \eqref{s_T}. 

Second, we check the conditions \eqref{WKB} and \eqref{N}. We can use $\lambda(r)\sim\f{\hbar}{\e(r)}$, \eqref{e=T} and \eqref{T_loc_rad} 
to get $\lambda(r)\sim n_f^{\f{1}{4}}\sqrt{l_p r}$, 
which is shorter than $\MR(r)^{-\f{1}{2}}\sim r$ (from \eqref{rho_rad}) for $r\gg l_p$, 
and the condition \eqref{WKB} holds. 
Then, from \eqref{WKB} and \eqref{SB2}, we can estimate \eqref{N} as 
\begin{equation}\lb{N_rad}
n_f^{\f{1}{2}}\f{r}{l_p} \lesssim N(r) \lesssim n_f^{\frac{1}{4}} \frac{r^{\frac{3}{2}}}{l_p^{\frac{3}{2}}},
\end{equation}
which is large for $r\gg l_p$. 

Third, we study the relation to the second law. 
Locally $\f{\p\tilde s(r)}{\p T_{loc}(r)}>0$ holds from \eqref{SB2}, while for the whole part, we have $\f{\p S}{\p T_0}<0$ indicating that the total specific heat is negative \ci{Landau_SM}, where we use \eqref{S_total_rad} and $T_0=T_{loc}(R) \propto R^{-1/2}$ (from \eqref{T_loc_rad}). Thus, due to the long-range nature of gravity, the thermodynamic behavior is different depending on the region being considered. 

Fourth, we check that thermal radiation does not saturate the sufficient conditions for the Bousso bound.
We first study \eqref{no_trap_Bek}. 
The right hand side is evaluated as 
$ \f{1}{\hbar}\sqrt{g_{rr}(r)}(r-a(r)) 4\pi r^2 \bra -T^t{}_t(r)\ket \sim \f{r}{l_p^2}$ from \eqref{rho_rad} and \eqref{metricA3}, 
while the left hand side as $s=4\pi r^2 s_{3d} \sim n_f^{\f{1}{4}}\f{r^{\f{1}{2}}}{l_p^{\f{3}{2}}}$ from \eqref{SB2}. 
Therefore, \eqref{no_trap_Bek} is not saturated for $r\gg l_p$ and not so large $n_f$. 

We next examine \eqref{BFM_cond}. 
The right-hand side becomes 
$\rho+p=\f{4}{3}\rho=\f{1}{14\pi G r^2}$ from \eqref{rho_rad}. 
For the left hand side, 
we have $\a(r)=\f{\p_r \log \sqrt{-g_{tt}(r)}}{\sqrt{g_{rr}(r)}}\sim \f{1}{r}$ from \eqref{metricA3}  
and obtain from \eqref{SB2} 
$\l|\f{\hbar \a}{2\pi} \f{s}{4\pi r^2} - \f{\hbar}{2\pi} \f{1}{\sqrt{g_{rr}}}\p_r \l(\f{s}{4\pi r^2}\r)\r|
\sim \f{n_f^{\f{1}{4}}}{Gr^2}\sqrt{\f{l_p}{r}}$. 
Thus, \eqref{BFM_cond} is not saturated. 

\section{Self-consistency of $(g_{\mu\nu}^*,|\psi\ket_*)$}\label{A:EMT} 
We give a short review for the derivation of the self-consistent values \eqref{sigma} of $(\s,\eta)$, where the origin of the tangential pressure can be seen (see Ref.\ci{KY4} for details). This is also a demonstration of our self-consistent treatment in the semi-classical Einstein equation \eqref{Einstein}.

We start with a review about how to solve the semi-classical Einstein equation \eqref{Einstein} self-consistently.
First, we consider the physical system and problem of interest 
and construct a candidate metric $g_{\mu\nu}$ for it, say, by making a model or a thought experiment.
Next, using that metric as a background spacetime, 
we examine the behavior of the matter fields and identify a candidate state $|\psi\ket$ for the system. 
Then, we use the solutions of the matter field equations, construct the regularized energy-momentum tensor, and renormalize it to remove the UV divergences. Finally, we equate the obtained renormalized energy-momentum tensor $\bra\psi|T_{\mu\nu}|\psi\ket$ 
with the Einstein tensor $G_{\mu\nu}$ calculated from the metric and solve \eqref{Einstein}. 
If it can be solved consistently, the candidate  $(g_{\mu\nu},|\psi\ket)$ is the self-consistent solution. If not, we set up another candidate and repeat the procedure. Note that this self-consistent analysis of \eqref{Einstein} allows us to obtain a non-perturbative solution in $\hbar$. 

Let us implement this program by setting  \eqref{interior} for the inner region \eqref{a*_range} and the Schwarzschild metric with $a_0$ for the outer one $R\leq r$ as the candidate metric. 

\textbf{(1) Candidate state.} 
It should be natural to consider its formation process and find a candidate state for the gravity condensate. As discussed in Sec.\ref{s:formation}, we can form it slowly in a heat bath of Hawking temperature $\sim \f{\hbar}{a_0}$. Studying each mode of scalar field, $\phi(x)=e^{-i \o t}\varphi(r)Y_{lm}(\theta,\phi)$, in the formation, we can find that there exist two types of modes inside. One is bound modes with various angular momenta $l$, which are trapped inside the condensate and cannot be excited due to a constraint from the Bohr-Sommerfeld quantization condition. The other is a continuum mode of s-waves, which can go to and from the outside and can be excited with $\o\sim \f{1}{a_0}$ to express thermal radiation from the bath and the Hawking radiation produced inside. Therefore, we set a candidate state $|\psi\ket_*$ such that 
\begin{equation}\lb{psi_EMT}
\bra \psi| T_{\mu\nu} | \psi \ket_* \approx \bra 0| T_{\mu\nu} | 0 \ket+T_{\mu\nu}^{(\psi)}.
\end{equation}
The first term is the contribution from the vacuum fluctuation of all modes in the ground state $|0\ket$ in \eqref{interior}, and the second one is that from the excitation of the continuum mode of s-waves. We here assume that the second contribution is so excited that it can be approximated as a classical one. 

\textbf{(2) Regularization.} 
To regularize the energy-momentum operator, we use the dimensional regularization scheme. 
In a $d=4+\e$ dimensional spacetime, the semi-classical Einstein equation 
coupled with $n_s$ free massless scalar fields can be expressed 
as a regularized operator equation \ci{BD}: 
\begin{align}\lb{geq1}
&G_{\mu\nu}=8\pi G \l[\mu^{-\e}T_{\mu\nu}-2\l(\f{\hbar n_s}{1152\pi^2\e}+\a(\mu)\r)H_{\mu\nu}\r.\nn\\
  &\l.-2 \l(-\f{\hbar n_s}{2880\pi^2\e}+\b(\mu)\r)K_{\mu\nu}
  -2 \l(\f{\hbar n_s}{2880\pi^2\e}+\g(\mu)\r)J_{\mu\nu}
  \r].
\end{align}
Here, $\hbar \mu$ is a renormalization point, 
$T_{\mu\nu}$ is the regularized energy-momentum tensor operator, 
and the other tensors are proportional to the identity operator.
The counter terms with $1/\e$ is chosen as those required by the minimal subtraction scheme, 
and each tensor is defined as, respectively, 
$H_{\mu\nu} \equiv \f{1}{\sqrt{-g}}\f{\d}{\d g^{\mu\nu}}\int d^dx\sqrt{-g}R^2$, 
$K_{\mu\nu} \equiv \f{1}{\sqrt{-g}}\f{\d}{\d g^{\mu\nu}}\int d^dx\sqrt{-g}R_{\a\b}R^{\a\b}$, 
and 
$J_{\mu\nu} \equiv \f{1}{\sqrt{-g}}\f{\d}{\d g^{\mu\nu}}\int d^nx\sqrt{-g}R_{\a\b\g\d}R^{\a\b\g\d}$. 
We also have the renormalized coupling constants at energy scale $\hbar\mu$: 
\begin{align}
\a(\mu)&=\a_0-\f{\hbar n_s}{2304\pi^2}\log\l(\f{\mu^2}{\mu_0^2}\r),~\nn\\
\b(\mu)&=\b_0+\f{\hbar n_s}{5760\pi^2}\log\l(\f{\mu^2}{\mu_0^2}\r),~\nn\\
\g(\mu)&=-\f{\hbar n_s}{5760\pi^2}\log\l(\f{\mu^2}{\mu_0^2}\r).
\end{align}
Here, $\a_0$ and $\b_0$ fix a 4D theory at energy scale $\hbar\mu_0$ 
while we have chosen $\g_0=0$ because of the 4D Gauss-Bonnet theorem. 
Thus, we obtain a more explicit form of the renormalized energy-momentum tensor in the right-hand side of \eqref{Einstein}: 
\begin{align}\label{EMT_ren}
&\bra \psi|T_{\mu\nu}|\psi \ket= \lim_{\e\to0}\l[\mu^{-\e}\bra \psi|T_{\mu\nu}|\psi \ket_{reg}\r.\nn\\
&-\f{\hbar n_s}{1440\pi^2\e}\l(\f{5}{2}H_{\mu\nu}-K_{\mu\nu}+J_{\mu\nu}\r)\nn\\
&\l.-2 (\a(\mu)H_{\mu\nu}+\b(\mu)K_{\mu\nu}+\g(\mu)J_{\mu\nu})\r].
\end{align}

\textbf{(3) Matter fields.} 
We now focus on $\bra 0| T_{\mu\nu} | 0 \ket$ in \eqref{psi_EMT}, 
which will lead to the self-consistent value of $\s$. 
To evaluate $\bra 0|T_{\mu\nu}|0 \ket_{reg}$, 
we solve the matter field equation $\Box \phi=0$ in 
the $(4+\e)$-dimensional spacetime manifold $\MM\times \mathbb{R}^\e$, 
where $\MM$ is our 4D physical spacetime \eqref{interior} and $\mathbb{R}^\e$ is $\e$-dimensional flat spacetime \ci{KN}:
\begin{equation}\lb{met_reg}
ds^2=-\f{2\s}{r^2} e^{\f{r^2}{2\s \eta}} d t^2 + \f{r^2}{2\s} d r^2 + r^2 d \Omega^2+\sum_{a=1}^\e (dy^a)^2.
\end{equation}
Using the fact that \eqref{interior} is locally $AdS_2\times S^2$ (with AdS radius $L\equiv \sqrt{2\eta^2\s}$ and $S^2$ radius $r$ as in footnote \ref{foot:AdS}), 
we can solve $\Box \phi=0$ around a point $r=r_0$ 
perturbatively by a $1/r$ expansion for $r\gg L$.
Then, the 0-th order solution for the bound modes is given by 
\begin{equation}\lb{phireg}
\phi(x)=\sum_i (a_iu_i^{(0)}(x)+a_i^\dagger u_i^{(0)\ast}(x)),
\end{equation}
where $a_i|0\ket=0$, $\sum_i=\sum_{n,l,m}\int d^\e k$, and
\begin{align}
&u_i^{(0)}(t,r,\th,\phi,y^a)= \nn\\
&\sqrt{\f{\hbar\eta}{2r_0}}\sqrt{\f{\p \o_{nl}}{\p n}} e^{-i\o_{nl} t}e^{-\f{r^2}{8\s\eta}}J_j(X) Y_{lm}(\th,\phi) \f{e^{ik\cdot y}}{(2\pi)^{\e/2}}.
\end{align}
Here, $n$ represents the quantum number satisfying the Bohr-Sommerfeld quantization condition; the factors in front of $e^{-i\o_{nl} t}$ is the normalization consistent with the condition and commutation relation; and $J_j(X)$ is the Bessel function, where $j \equiv \sqrt{L^2 (l(l+1)/r_0^2 +k^2) +\f{1}{4}}$ 
and $X\equiv L e^{\f{r_0(r_0-r)}{2\s\eta}}\o/\sqrt{-g_{tt}(r_0)}$. 

\textbf{(4) Renormalization.} 
After a long calculation using \eqref{phireg}, we can obtain the leading term of the renormalized energy-momentum tensor \eqref{EMT_ren}: 
\begin{align}\lb{T0ren}
\bra0|T^\mu{}_\nu|0\ket^{(0)}
=&\left(
\begin{array}{cccc}
1 & & &  \\
 & 1& &  \\
 & & -1&  \\
 & & & -1
\end{array}
\right)\bra0|T^t{}_t|0\ket^{(0)}\nn\\
&+
\left(
\begin{array}{cccc}
0 & & &  \\
 &0 & &  \\
 & & 1&  \\
 & & &1 
\end{array}
\right)\f{\hbar n_s}{1920\pi^2\eta^4\s^2},
\end{align}
where the components are in the order of $(t,r,\theta,\phi)$ and
\begin{align}\lb{T0tren}
&\bra0|T^t{}_t|0\ket^{(0)}\nn\\
&=\f{\hbar n_s}{1920\pi^2\eta^4\s^2}\l[2c+\g + \log\f{1}{32\pi \eta^2\s \mu^2_0}-\f{960\pi^2}{\hbar n_s }(2\a_0+\b_0) \r].
\end{align}
Here, $\g$ is Euler's constant and $c$ is a non-trivial finite value for $|0\ket$: $c=0.055868$. 
We note that the trace part
\begin{equation}\lb{trace0}
\bra0|T^\mu{}_\mu|0\ket^{(0)}=\f{\hbar n_s}{960\pi^2\eta^4\s^2}
\end{equation}
is independent of $(\a_0,\b_0)$. We can use \eqref{RRR} and check that this agrees with the leading value of the 4D Weyl anomaly \ci{BD,Nicolai}:
\begin{equation}\label{T_ano}
\bra T^\mu{}_\mu \ket_{Anomaly} = \frac{\hbar n_s}{2880 \pi^2}\left(R_{\mu\nu\alpha\beta}^2 -R_{\mu\nu}^2+\frac{5}{2}R_{RicciScalar}^2 \right).    
\end{equation}

\textbf{(5) Self-consistent solution.} 
Finally, we use \eqref{RRR}, \eqref{psi_EMT} and \eqref{trace0} and solve the trace part of \eqref{Einstein} at the leading order: 
\begin{align}\label{sigma1}
(G^\mu{}_\mu)^{(0)} &= 8\pi G \bra\psi|T^\mu{}_\mu|\psi\ket^{(0)} \nn\\
\Rightarrow \f{1}{\s\eta^2} &= 8\pi G \f{\hbar n_s}{960\pi^2\eta^4\s^2} \nonumber \\
\Rightarrow \s &= \f{n_s l_p^2}{120\pi \eta^2},
\end{align}
which gives the self-consistent value of $\s$ in \eqref{sigma}. 
Here, we have dropped the contribution from $T^{(\psi)\mu}{}_\mu$ 
because the mode integrations over $\o$ \textit{and} $l$ lead to $\MO(1)$ terms in \eqref{T0ren}, 
and $T^{(\psi)\mu}{}_\mu$, including only s-waves, cannot produce a $\MO(1)$ term. Note that \eqref{sigma1} is a non-perturbative argument in that it does not hold in the limit $\hbar \to 0$. 

Furthermore, we can consider $T_{\mu\nu}^{(\psi)}=\MO(r^{-2})$ 
and calculate the 1st order contribution $\bra0|T^\mu{}_\mu|0\ket^{(1)}=\MO(r^{-2})$ to find the self-consistent value of $\eta$ in a similar manner. 
Here, we need to choose a theory with $(\a_0,\b_0)$ such that $\bra0|T^t{}_t|0\ket^{(0)}=0$ in \eqref{T0tren} and $1\leq \eta <2$ in \eqref{sigma} hold. We thus conclude that $(g_{\mu\nu}^*,|\psi\ket_*)$ with \eqref{sigma} satisfies the semi-classical Einstein equation \eqref{Einstein} self-consistently and non-perturbatively in $\hbar$. 

Here, we can see explicitly the origin of the large tangential pressure $\bra\psi|T^\theta{}_\theta|\psi\ket^*$ in \eqref{EMT}. 
Under the condition that $\bra0|T^t{}_t|0\ket^{(0)}=0$, the second term in \eqref{T0ren} gives 
\begin{equation}
\bra0|T^\theta{}_\theta|0\ket^{(0)}=\f{\hbar n_s}{1920\pi^2\eta^4\s^2}=\f{15}{2G n_s l_p^2},
\end{equation}
where we have used \eqref{sigma1}. 
This is the half of the Weyl anomaly \eqref{trace0}, which is given by the curvatures \eqref{T_ano}. 
Therefore, the pressure originates from 
4D quantum fluctuations induced by the curved spacetime \eqref{interior}.

\section{Maximum entropy from uniformity}\label{A:N(r)}
We provide another derivation of \eqref{a_*}: we derive it from a radially-uniform condition, 
instead of setting $\e=\e_{max}$.  
This indicates that the radial uniformity is a sufficient condition for entropy maximization; since the converse proposition of this is given in Sec.\ref{s:a*}, we can therefore conclude that \textit{a necessary and sufficient condition for entropy maximization is radial uniformity.} This would be natural for a spherically symmetric system in equilibrium according to thermodynamics in flat space, but it is non-trivial for a self-gravitating system and consistent with Appendix.\ref{A:Tolman}.

To express radial uniformity, we use the occupation number \eqref{N} and impose 
\begin{equation}\label{N0}
    N(r)={\rm const}.\equiv N_0~~{\rm for}~\Delta \hat r \sim \lambda(r),
    \end{equation}
where $N_0=\MO(1)\gg1$. This means maximum uniformity, in the sense that the number of excited quanta is constant even in subsystems with the smallest width. 

We first prepare a useful formula for analyzing \eqref{N0}. From the discussion below \eqref{N}, we have $N(r)\sim \frac{\Delta E_{loc}}{\epsilon}\sim \frac{4\pi r^2 \hbar \bra -T^t{}_t\ket}{\epsilon^2}$ for $\Delta \hat r \sim \lambda \sim \frac{\hbar}{\epsilon}$. Applying \eqref{Hami}, we obtain 
\begin{equation}\label{N_a}
    N(r)\sim \frac{m_p^2}{\epsilon(r)^2} \partial_r a(r).
\end{equation}

Setting \eqref{N0}, we then combine \eqref{N_a} and \eqref{eq_a*} to get
\begin{equation}\lb{A_aeq1}
N_0 l_p^2 \sim r(r-a_*(r))\p_r a_*(r).
\end{equation}
Here, $N(r)={\rm const.}$ means through \eqref{N_a} 
that $\e(r)>0$ and $\p_r a_*(r)>0$. 
Hence, we can solve $a_*=a_*(r)$ for $r$ to have $r=r(a)$ 
(here we write $a=a_*$ for simplicity) and express \eqref{A_aeq1} as 
\begin{equation}\lb{A_aeq2}
\f{dr(a)}{da}=\f{1}{\g}r(a)(r(a)-a),
\end{equation}
where $\g={\rm const.}=\MO(N_0 l_p^2)$.

The general solution to \eqref{A_aeq2} is given by 
\begin{equation}\lb{r(a)1}
r(a)=\f{e^{-x^2}}{C-\sqrt{\f{\pi}{2\g}} {\rm erf}(x) },
\end{equation}
where $C$ is an integration constant, $x\equiv \f{a}{\sqrt{2\g}}$ and ${\rm erf}(x)\equiv \f{2}{\sqrt{\pi}}\int^x_0 dy e^{-y^2}$.
We are now considering a configuration with $\f{a_0}{2G}\gg m_p$,
and we can focus on the asymptotic form of \eqref{r(a)1} for $a\gg l_p$, that is, $x\gg1$: 
\begin{equation}\lb{r(a)2}
r(a)\approx\f{e^{-x^2}}{\sqrt{\f{\pi}{2\g}}-C + e^{-x^2} \l[\f{1}{\sqrt{2\g}x}-\f{1}{2\sqrt{2\g}x^3}+\MO(x^{-5}) \r] }.
\end{equation}
If $\sqrt{\f{\pi}{2\g}}-C\neq 0$, 
we would have $r(a)\sim e^{-x^2}$ for $x\gg1$, 
but it is not consistent with $r,a\gg l_p$. 
Therefore, we must have $C= \sqrt{\f{\pi}{2\g}}$. 
Then, we obtain 
\begin{align}\lb{A_aeq3}
r(a) &\approx \sqrt{2\g}x + \sqrt{\f{\g}{2}}\f{1}{x} + \MO(x^{-3}) \nn \\
 &= a + \f{\g}{a} +\MO(a^{-3}),
\end{align}
which means for $a\gg l_p$ 
\begin{equation}\lb{A_aeq4}
a_*(r)\approx r-\f{\g}{r}.
\end{equation}

We now substitute \eqref{A_aeq4} into \eqref{N_a} to get
\begin{equation}
\e(r)\sim \f{m_p}{\sqrt{N_0}}.
\end{equation}
Comparing this to \eqref{e_max}, we can identify $N_0\sim n$, since at this stage, $n$ is an arbitrary large number with $\MO(1)$. Therefore, \eqref{A_aeq4} means \eqref{a_*}.

\section{Tolman's law does not lead to maximum entropy.}\label{A:Tolman}
We show that keeping Tolman's law under the semi-classical condition \eqref{e_max} and consistency with local thermodynamics does \textit{not} yield maximum entropy for a given surface area. This gives an explanation for the violation of Tolman's law in maximizing the entropy (discussed in Sec.\ref{s:T_loc}). Note that this is consistent with Refs.\cite{Green,Xia}, where isotropic fluid is assumed and quantities other than a surface area are fixed (see also footnote \ref{foot:thermal}).

Suppose that, instead of setting $\e(r)=\e_{max}$ at each $r$, we use Tolman's law $T_{loc}(r)\sqrt{-g_{tt}(r)}={\rm const}.$ consistent with the semi-classical condition \eqref{e_max} and saturate the entropy bound \eqref{S_B}. 
From \eqref{e=T}, we have 
\begin{equation}\label{e_Tolman}
\epsilon(r)=\frac{\sqrt{-g_{tt}(r_0)}}{\sqrt{-g_{tt}(r)}} \epsilon_{max},    
\end{equation}
where the maximum excitation $\epsilon_{max}\sim \frac{m_p}{\sqrt{n}}$ is reached at an innermost radius $r_0$. To maximize the entropy in the bound \eqref{S_B}, then \eqref{e_Tolman} should saturate the local condition \eqref{no_trap}: 
$\frac{\hbar}{\epsilon(r)}\sim \sqrt{g_{rr}(r)}(r-a(r))=\frac{r}{\sqrt{g_{rr}(r)}}$.  
Taking the square of the both hands and applying \eqref{e_Tolman}, we get 
\begin{equation}\label{e_Tolman2}
    (-g_{tt}(r))g_{rr}(r)=C r^2,
\end{equation}
where $C$ is a positive constant. Therefore, the interior metric is given by 
\begin{equation}\label{Tolman_metric}
    ds^2=-\frac{Cr^2}{g_{rr}(r)}dt^2+g_{rr}(r)dr^2+r^2d\Omega^2.
\end{equation}

To examine the asymptotic form of $g_{rr}(r)$ for large $r$, we set $g_{rr}(r)=B_0 r^k$, where $B_0$ is positive because of the condition \eqref{no_trap}, 
and study the Einstein tensors: 
\begin{align}
    -G^t{}_t(r)&=\frac{1}{r^2}+\frac{k-1}{B_0}r^{-2-k},\\
    G^r{}_r(r)&=-\frac{1}{r^2}+\frac{3-k}{B_0}r^{-2-k},\\
    G^\theta{}_\theta(r)&=\frac{k^2-4k+2}{2B_0}r^{-2-k}.
\end{align}
As done in Sec.\ref{s:g*}, we employ consistency with local thermodynamics (i.e. positivity of energy density and pressures). 
If $k<0$, $-G^t{}_t(r)$ would be negative for large $r$, meaning a negative energy density. Therefore, we get a condition $k\geq 0$. In order for the radial pressure to be positive for large $r$, we must have $r^{-2-k}\geq r^{-2}$ and $3-k>0$, leading to $k\leq 0$. Thus, we conclude $k=0$. 

Then, the metric \eqref{Tolman_metric} reduces to 
\begin{equation}\label{Tolman_metric2}
    ds^2=-\frac{C}{B_0}r^2dt^2+B_0 dr^2+r^2d\Omega^2,
\end{equation}
and the Einstein tensors become 
\begin{equation}\label{G_B0}
    -G^t{}_t(r)=\frac{B_0-1}{B_0}r^{-2},~G^r{}_r(r)=\frac{3-B_0}{B_0}r^{-2},~G^\theta{}_\theta(r)=\frac{1}{B_0}r^{-2}.
\end{equation}
For these to be positive, we must have 
\begin{equation}\label{Tolman_metric3}
    1<B_0<3.
\end{equation}
Then, the energy distribution is given from \eqref{Hami} by 
\begin{equation}\label{a_B0}
    a(r)=\frac{B_0-1}{B_0} r < \frac{2}{3} r,
\end{equation}
which is smaller than the maximum one \eqref{a_*} (and Buchdahl's limit, $\frac{8}{9}r$ \cite{Buchdahl}). Through the right-hand side of \eqref{S_B}, therefore, the entropy of the configuration \eqref{Tolman_metric2} is $\sim \frac{B_0-1}{B_0}\frac{R^2}{l_p^2}< \frac{2}{3}\frac{R^2}{l_p^2}$. This cannot exceed $\sim \frac{R^2}{l_p^2}$ \eqref{S_*}, the value estimated for the gravity condensate in the same way. Thus, we conclude that Tolman's law does not maximize entropy for a given surface area.


Finally, we discuss an interesting point emerging as a by-product. For $B_0=2$, \eqref{G_B0} gives $-G^t{}_t=G^r{}_r=G^\theta{}_\theta=\frac{1}{2r^2}$. This represents Zel’dovich's causal-limit fluid, where $\rho=p$ holds for $\rho=\bra-T^t{}_t\ket,~p=\bra T^r{}_r\ket=\bra T^\theta{}_\theta \ket$ \cite{Zeldovich}. The entropy $S \propto \frac{R^2}{l_p^2}$ can be obtained directly by applying the Gibbs relation $T_{loc} s_{3d}=\rho+p$, Tolman's law $T_{loc}(r)=\frac{T_0}{\sqrt{-g_{tt}(r)}}$ and the metric \eqref{Tolman_metric2} to the formula \eqref{S}. This is a result supporting the consistency of our typicality argument. Also, the area-scaling entropy is consistent with Ref.\cite{Banks}. It would be interesting to study the relation between their derivation and ours. 



\begin{thebibliography}{99}
\bibitem{LIGO}
B.~P.~Abbott \textit{et al.} [LIGO Scientific and Virgo],
Phys. Rev. Lett. \textbf{116}, no.6, 061102 (2016).

\bibitem{EHT}
K.~Akiyama \textit{et al.} [Event Horizon Telescope],
Astrophys. J. Lett. \textbf{875}, L1 (2019).

\bibitem{Cardoso}
V.~Cardoso and P.~Pani,
Living Rev.Rel.\textbf{22},no.1,4(2019).

\bibitem{Bekenstein} 
J.~D.~Bekenstein,
Phys.\ Rev.\ D {\bf 7}, 2333 (1973); Phys.\ Rev.\ D {\bf 9}, 3292 (1974).

\bibitem{Hawking} 
  S.~W.~Hawking,
  Commun.\ Math.\ Phys.\  {\bf 43}, 199 (1975)  [Erratum-ibid.\  {\bf 46}, 206 (1976)].  

\bibitem{Dvali3}
G.~Dvali and C.~Gomez,
Fortsch. Phys. \textbf{61}, 742-767 (2013).


\bibitem{Oriti1}
D.~Oriti, D.~Pranzetti and L.~Sindoni,
Phys. Rev. Lett. \textbf{116}, no.21, 211301 (2016).


\bibitem{Bousso1}
R.~Bousso,
JHEP \textbf{07}, 004 (1999).


\bibitem{Bousso2}
R.~Bousso,
Rev. Mod. Phys. \textbf{74}, 825-874 (2002). 

\bibitem{Laurent}
W.~Donnelly and L.~Freidel,
JHEP \textbf{09}, 102 (2016). 


\bibitem{Landau_SM}
L. D. Landau and E. M. Lifshitz, \textit{Statistical Physics} (Butterworth-Heinemann, Oxford, 1984).

\bibitem{Pad_thermo}
T.~Padmanabhan,
Phys. Rept. \textbf{188}, 285 (1990).



\bibitem{Tolman}
R.~Tolman and P.~Ehrenfest,
Phys. Rev. \textbf{36}, no.12, 1791-1798 (1930).

\bibitem{Sorkin}
R.~D.~Sorkin, R.~M.~Wald and Z.~J.~Zhang,
Gen. Rel. Grav. \textbf{13}, 1127-1146 (1981). 


\bibitem{BD}
N.~D.~Birrell and P.~C.~W.~Davies, \textit{Quantum Fields in Curved space} 
(Cambridge University Press, 1982). 

\bibitem{Kiefer}
C.~Kiefer, \textit{Quantum Gravity} (Oxford University Press, Oxford, 2012). 

\bibitem{Zubarev}
D.N.Zubarev, A.V.Prozorkevich, and S.A.Smolyanskii,
Theoret. and Math. Phys. 40, 821 (1979).

\bibitem{FMW}
E.~E.~Flanagan, D.~Marolf and R.~M.~Wald,
Phys. Rev. D \textbf{62}, 084035 (2000).

\bibitem{BFM}
R.~Bousso, E.~E.~Flanagan and D.~Marolf,
Phys. Rev. D \textbf{68}, 064001 (2003).

\bibitem{Francesco}
F.~Becattini and D.~Rindori,
Phys. Rev. D \textbf{99}, no.12, 125011 (2019)

\bibitem{Goldstein}
S.~Goldstein, J.~L.~Lebowitz, R.~Tumulka and N.~Zanghi,
Phys. Rev. Lett. \textbf{96}, 050403 (2006).

\bibitem{Popescu}
S. Popescu, A. J. Short, and A. Winter, 
Nat. Phys. 2, 754 (2006).

\bibitem{Sugita}
A. Sugita, 
Nonlinear Phenom. Complex Sys. 10, 192 (2007).

\bibitem{Reimann}
P. Reimann, 
Phys. Rev. Lett. 99, 160404 (2007).

\bibitem{Bek_bound}
J.~D.~Bekenstein,
Phys. Rev. D \textbf{23}, 287 (1981).

\bibitem{Dvali1}
G.~Dvali, 
Fortsch. Phys. \textbf{58}, 528-536 (2010).

\bibitem{Dvali2}
G.~Dvali and M.~Redi,
Phys. Rev. D \textbf{77}, 045027 (2008).


\bibitem{Buchdahl}
H.~A.~Buchdahl,
Phys. Rev. \textbf{116}, 1027 (1959).

\bibitem{CY}
C.~Y.~Chen and Y.~Yokokura,
Phys. Rev. D \textbf{109}, no.10, 104058 (2024).


\bibitem{KMY} 
H.~Kawai, Y.~Matsuo, and Y.~Yokokura,
Int.\ J.\ Mod.\ Phys.\ A {\bf 28}, 1350050 (2013).  

\bibitem{KY1}
  H.~Kawai and Y.~Yokokura, 
  Int.\ J.\ Mod.\ Phys.\ A {\bf 30}, 1550091 (2015).  

\bibitem{KY4}
H.~Kawai and Y.~Yokokura,
Universe \textbf{6}, no.6, 77 (2020). 

\bibitem{KY5}
H.~Kawai and Y.~Yokokura,
Phys. Rev. D \textbf{105}, no.4, 045017 (2022).

\bibitem{Y1}
Y.~Yokokura,
Nucl. Phys. B \textbf{1002}, 116531 (2024).

\bibitem{BY1}
J.~D.~Brown and J.~W.~York, Jr.,
Phys. Rev. D \textbf{47}, 1420-1431 (1993).

\bibitem{BY2}
J.~D.~Brown and J.~W.~York, Jr.,
[arXiv:gr-qc/9301018 [gr-qc]].

\bibitem{Jacobson_vol}
T.~Jacobson and M.~R.~Visser,
Phys. Rev. Lett. \textbf{130}, no.22, 221501 (2023). 

\bibitem{Minic}
D.~Marolf, D.~Minic and S.~F.~Ross,
Phys. Rev. D \textbf{69}, 064006 (2004).

\bibitem{Casini}
H.~Casini,
Class. Quant. Grav. \textbf{25}, 205021 (2008).

\bibitem{Jacobson_entangle}
T.~Jacobson,
Phys. Rev. Lett. \textbf{116}, no.20, 201101 (2016).

\bibitem{KY2}
H.~Kawai and Y.~Yokokura,
Phys. Rev. D \textbf{93}, no.4, 044011 (2016).

\bibitem{Jacobson}
T.~Jacobson,
Phys. Rev. Lett. \textbf{75}, 1260-1263 (1995).

\bibitem{Pad}
T.~Padmanabhan,
Rept. Prog. Phys. \textbf{73}, 046901 (2010).


\bibitem{echo}
J.~Abedi, H.~Dykaar and N.~Afshordi,
Phys. Rev. D \textbf{96}, no.8, 082004 (2017).

\bibitem{Hayward}
S.~A.~Hayward,
Phys. Rev. D \textbf{53}, 1938-1949 (1996).

\bibitem{fluidbook}
P.~Romatschke and U.~Romatschke,
\textit{Relativistic Fluid Dynamics In and Out of Equilibrium}
(Cambridge University Press, Cambridge, 2019).

\bibitem{BCFM}
R.~Bousso, H.~Casini, Z.~Fisher and J.~Maldacena,
Phys. Rev. D \textbf{90}, no.4, 044002 (2014).


\bibitem{Caianiello}
E.~R.~Caianiello and G.~Landi,
Lett. Nuovo Cim. \textbf{42}, 70 (1985)

\bibitem{Pad_Lim}
T.~Padmanabhan,
Class. Quant. Grav. \textbf{4}, L107-L113 (1987).

\bibitem{Brandt}
H.~E.~Brandt,
Found. Phys. Lett. \textbf{2} (1989), 39.


\bibitem{Antonov}
V.~A.~Antonov, Vest. Leningrad Univ. \textbf{7}, 135 (1962). 

\bibitem{Lynden}
D.~Lynden-Bell and R.~Wood,
Mon. Not. Roy. Astron. Soc. \textbf{138}, 495 (1968).

\bibitem{Green}
S.~R.~Green, J.~S.~Schiffrin and R.~M.~Wald,
Class. Quant. Grav. \textbf{31}, 035023 (2014).

\bibitem{Xia}
M.~Xia and S.~Gao,
Eur. Phys. J. Plus \textbf{139}, no.6, 500 (2024).


\bibitem{No_trap}
M.~Mars and J.~M.~M.~Senovilla,
Class. Quant. Grav. \textbf{20}, L293-L300 (2003).


\bibitem{HKLY}
P.~M.~Ho, H.~Kawai, H.~Liao and Y.~Yokokura,
Eur. Phys. J. C \textbf{84}, 711 (2024). 

\bibitem{KY3}
H.~Kawai and Y.~Yokokura,
Universe \textbf{3}, no.2, 51 (2017)

\bibitem{Poisson}
E. Poisson, \textit{A Relativistic Toolkit} (Cambridge University Press, Cambridge, 2004).

\bibitem{Barcelo}
C.~Barcelo, S.~Liberati, S.~Sonego and M.~Visser,
Phys. Rev. D \textbf{83}, 041501 (2011). 


\bibitem{Oppenheim}
J.~Oppenheim,
Phys. Rev. D \textbf{65}, 024020 (2002); Phys. Rev. E \textbf{68}, 016108 (2003). 

\bibitem{Groot}
S. R. de Groot and P. Mazur, \textit{Non-Equilibrium Thermodynamics} (North-Holland, Amsterdam, 1962).

\bibitem{GH}
G.~W.~Gibbons and S.~W.~Hawking,
Phys. Rev. D \textbf{15}, 2752-2756 (1977).

\bibitem{Nakagawa-Sasa}
N.~Nakagawa and S.~i.~Sasa,
J. Stat. Phys. \textbf{177}, 825-888, (2019).


\bibitem{Pesci}
A.~Pesci,
Class. Quant. Grav. \textbf{24}, 6219-6226 (2007); 
Class. Quant. Grav. \textbf{25}, 125005 (2008).

\bibitem{Kuchar}
K.~Kuchar,
J. Math. Phys. \textbf{11}, 3322-3334 (1970). 

\bibitem{Suvrat}
C.~Chowdhury, V.~Godet, O.~Papadoulaki and S.~Raju,
JHEP \textbf{03}, 019 (2022). 

\bibitem{Landau_SM2}
E. M. Lifshitz and L. P. Pitaevskii, \textit{Statistical Physics (Part2): Theory of the Condensed State} (Butterworth-Heinemann, Oxford, 1980).



\bibitem{Bianca}
B.~Dittrich, T.~Jacobson and J.~Padua-Arg\"uelles,
Phys. Rev. D \textbf{110}, no.4, 046006 (2024).

\bibitem{BHmodel}
A.~Fabbri and J.~Navarrro-Salas, \textit{Modeling Black Hole Evaporation} (Imperial College Press, London, 2005).

\bibitem{B_singularity}
R.~Bousso and A.~Shahbazi-Moghaddam,
Phys. Rev. Lett. \textbf{128}, no.23, 231301 (2022).


\bibitem{Y3}
Y.~Yokokura, to appear. 

\bibitem{Wald}
R.~M.~Wald,
Phys. Rev. D \textbf{48} (1993) no.8, R3427-R3431.

\bibitem{SY}
S.~i.~Sasa and Y.~Yokokura,
Phys. Rev. Lett. \textbf{116}, no.14, 140601 (2016).

\bibitem{Weinberg}
S.~Weinberg, \textit{Gravitation and Cosmology} (Wiley, 1972). 


\bibitem{KN} 
  H.~Kawai and M.~Ninomiya,
  Nucl.\ Phys.\ B {\bf 336}, 115 (1990).

\bibitem{Nicolai}
L.~Casarin, H.~Godazgar and H.~Nicolai,
Phys. Lett. B \textbf{787}, 94-99 (2018).

\bibitem{Zeldovich}
Y.~B.~Zel'dovich,
Zh. Eksp. Teor. Fiz. \textbf{41}, 1609-1615 (1961).

\bibitem{Banks}
T.~Banks, W.~Fischler, A.~Kashani-Poor, R.~McNees and S.~Paban,
Class. Quant. Grav. \textbf{19}, 4717-4728 (2002). 

\end{thebibliography}
\end{document}